\documentclass[11pt,hyper,letter]{JHEP3}
\usepackage[dvips]{epsfig}
\usepackage{graphicx,amssymb,amsmath,amsfonts,cite}
\usepackage[english]{babel}
\newcommand{\cA}{{\cal A}}

  \newcommand{\cL}{{\cal L}}
  \newcommand{\cN}{{\cal N}}
\newcommand{\cO}{{\cal O}}

\newcommand{\cW}{{\cal W}}

\newcommand{\be}{\begin{equation}} \newcommand{\ee}{\end{equation}}
\newcommand{\bea}{\begin{eqnarray}} \newcommand{\eea}{\end{eqnarray}}
\newcommand{\beann}{\begin{eqnarray*}}  \newcommand{\eeann}{\end{eqnarray*}}
\newcommand{\bfig}{\begin{figure}} \newcommand{\efig}{\end{figure}}
\newcommand{\ba}{\begin{array}} \newcommand{\ea}{\end{array}}
\newcommand{\bcen}{\begin{center}} \newcommand{\ecen}{\end{center}}
\newcommand{\btab}{\begin{tabular}} \newcommand{\etab}{\end{tabular}}

\newcommand{\vev}[1]{\left\langle{#1}\right\rangle}

\newtheorem{Proposition}{Proposition}[section]

\newtheorem{Theorem}{Theorem}[section]
\newtheorem{Lemma}{Lemma}[section]

\newcommand{\bp}{\begin{Proposition}}   \newcommand{\ep}{\end{Proposition}}
\newcommand{\bt}{\begin{Theorem}}   \newcommand{\et}{\end{Theorem}}
\newcommand{\bl}{\begin{Lemma}}     \newcommand{\el}{\end{Lemma}}
\newcommand{\bc}{\begin{Corolary}} \newcommand{\ec}{\end{Corolary}}
\def\rhuv{{\rho_{UV}}}
\def\rhir{{\rho_{IR}}}
\def\Luv{{L_{UV}}}
\def\Lir{{L_{IR}}}

\title{\LARGE The $a$-theorem and conformal symmetry breaking in holographic RG flows}

\author{Carlos Hoyos, Uri Kol, Jacob Sonnenschein and Shimon Yankielowicz
\\
  Raymond and Beverly Sackler Faculty of Exact Sciences \\
School of Physics and Astronomy \\
Tel-Aviv University, Ramat-Aviv 69978, Israel.}

\footnotetext{E-mail addresses: choyos, urikol, cobi, shimonya @post.tau.ac.il}

\abstract{We study holographic models describing an RG flow between two fixed points driven by a relevant scalar operator. We show how to introduce a spurion field to restore Weyl invariance and compute the anomalous contribution to the generating functional in even dimensional theories. We find that the coefficient of the anomalous term is proportional to the difference of the conformal anomalies of the UV and IR fixed points, as expected from anomaly matching arguments in field theory. For any even dimensions the coefficient is positive as implied by the holographic $a$-theorem. For flows corresponding to spontaneous breaking of conformal invariance, we also compute the two-point functions of the energy-momentum tensor and the scalar operator and identify the dilaton mode. Surprisingly we find that in the simplest models with just one scalar field there is no dilaton pole in the two-point function of the scalar operator but a stronger singularity. We discuss the possible implications.}


\preprint{TAUP-2953/12}

\begin{document}

\section{Introduction and summary}

Recently Komargodski and Schwimmer have proved  the four dimensional ``$a$-theorem"  \cite{Komargodski:2011vj,Komargodski:2011xv}.
This theorem, conjectured originally by Cardy \cite{Cardy:1988cwa}, states that for  any Renormalization Group (RG) flow between an UV and an IR  fixed points  the following inequality holds
\be
a_{UV} \geq a_{IR}
\ee
where $a_{UV}$ and $ a_{IR}$ are the UV and IR conformal anomalies related to the expectation value of the energy-momentum tensor as follows
\begin{equation}\label{Afour}
\vev{T^\mu_\mu}=c \cW^2-a E_4,
\end{equation}
 where $\cW^2$ is the Weyl tensor squared and $E_4$ is the Euler density. The $a$-theorem is the four dimensional relative of the well known two dimensional Zamolodchikov's  $c$-theorem \cite{Zamolodchikov:1986gt}. The proof of the $a$-theorem of \cite{Komargodski:2011vj}
 follows from the analyticity and unitarity of the forward 4-point scattering amplitude
of a scalar field $\tau(x)$. The latter corresponds to the dilaton, the Goldstone boson associated with the spontaneously broken conformal invariance or to  the conformal compensator (spurion) for RG flows driven by relevant operators.
Another interesting result based on the ideas behind the $a$-theorem is a proof that perturbative theories always flow to CFTs \cite{Luty:2012ww}, although it seems that some aspects remain to be understood \cite{Fortin:2012cq}.

Since there is a monotonically decreasing function associated with  RG flows in two and four dimensions, a natural question to raise  is whether this theorem can be generalized to  field theories at any even dimensions.
It turns out that the generalization of the arguments of \cite{Komargodski:2011vj}  to six dimensional field theories   is not straightforward as was revealed in \cite{Elvang:2012st}.
An alternative to proving a generalized theorem directly for $d$ dimensional  field theory is to do it using  a dual    $d+1$ dimensional holographic  gravitational bulk  system.

RG flows have been  intensively studied   using holography, starting with the pioneering works  \cite{Akhmedov:1998vf,Balasubramanian:1999jd,de Boer:1999xf} and soon later within the formalism of holographic renormalization \cite{Bianchi:2001kw,Bianchi:2001de}.  In most of the cases
 (e.g.\cite{Freedman:1999gk,Freedman:1999gp,Girardello:1999bd,Arutyunov:2000rq,Bigazzi:2001ta,Berg:2001ty,Halmagyi:2005pn,Hotta:2008xt}), the RG flows  are driven by a relevant deformation of the Lagrangian. RG flows associated with spontaneous breaking of conformal symmetry were studied in,  for instance, \cite{Martelli:2001tu,D'Hoker:2002aw,Freedman:2003ax}. Based on holography, $a$-theorems in arbitrary number of  dimensions where proven using various different approaches. By studying the renormalization group flow along null geodesic congruences the authors of  \cite{Alvarez:1998wr} have proven  a holographic version  of an  ``$a$-theorem".
 An ``$a$-function" that is positive and monotonic if a weak energy condition holds in the bulk gravity theory was constructed  in \cite{Freedman:1999gp}. For even-dimensional boundaries, the $a$-function  was shown to coincide  with the trace anomaly coefficients  of  the corresponding  boundary  field theory in limits where conformal invariance is recovered.
 In \cite{Myers:2010tj} it was demonstrated that  unitary higher curvature gravity theories in both odd and even  dimensions admit    $a$-theorems in terms of their corresponding  entanglement entropy.   In cases where the dual CFT is even-dimensional it was shown   that  it is the coefficient of the ``$a$-anomaly"  that flows.

The RG flows we study have a holographic dual description as a solitonic solution interpolating between two $AdS$ spaces of dynamical Einstein gravity coupled to a single scalar field in $(d+1)$ dimensions, which is generated by an appropriate scalar potential. We determine the solution of the scalar field and of  the  metric that admits a  $d$-dimensional Poincar\'e isometry. The scalar and the metric both   depend  only on the $(d+1)^{th}$ dimensional (radial) coordinate. We then  introduce the Fefferman-Graham coordinates to express the general fluctuations of the metric components and  of the scalar field in the $AdS$ regions in terms of a power expansion in the radial coordinate which is also an expansion in powers of the  $d$ dimensional  space-time derivatives. This expansion is valid up to some value of the radial coordinate, so we need to introduce a cutoff in the radial direction that acts as an IR regulator. We then introduce the spurion via a generalization of the Penrose-Brown-Henneaux (PBH) diffeomorphism \cite{Penrose:1986ca,Brown:1986nw}, whose effect is to perform a Weyl transformation on the boundary values of the metric and the scalar. The same kind of transformation (in pure $AdS$) has been used recently to compute the effective action of the dilaton from the action of probe branes \cite{Elvang:2012st}. A generalization of the transformation to theories with scalars was also used in \cite{Hung:2011ta} to compute contributions of relevant operators to the entanglement entropy. Let us stress here that in our case there is no analog to the scalar fields on the probe branes and the parameter of the PBH transformation is not a dynamical mode.\footnote{However, it could be promoted to a dynamical mode by changing the kind of boundary conditions of the metric, we discuss this possibility in Sec.~\ref{sec:promote}}

The PBH transformation admits the same kind of expansion as the metric and scalar fluctuations, in such a way that terms in the expansion transform covariantly. There is an exception that appears at $d$th order in derivatives, whose effect is to introduce a term related to the Weyl anomaly in the generating functional computed using the usual holographic prescription. The term takes the  following form
 \begin{equation}
 \int d^d x\,\sqrt{-\hat g}\,\tau(x)\,\left(\hat \cA_d^{UV}-\hat \cA_d^{IR}\right),
\end{equation}
where $\tau(x)$ is the spurion field and $\hat g$ is the determinant of a background metric, dressed with the spurion to make a Weyl-invariant combination, i.e. $\hat g_{\mu\nu}=e^{2\tau}g_{\mu\nu}$. $\cA_d$ is the conformal anomaly in even dimensions, at the UV or IR fixed points. In general, the anomaly term takes the form \cite{Boulanger:2007ab}
\begin{equation}
\cA_d=-a E_d+\sum  c_i I_i,
\end{equation}
where $E_d$ is the Euler density and $I_i$ are Weyl-invariant polynomials of the curvature and its derivatives with $d$ derivatives of the metric, examples in $d=2$  and $d=4$ are given by \eqref{Atwo},\eqref{Afour2}. When the action of the gravitational dual is Einstein gravity plus matter fields, as in our case, the value of the coefficients $c_i$ are proportional to $a$ with fixed factors. In more general cases than the ones we consider the coefficients $c_i$ are independent. Our computation confirms that there is a term in the generating functional of the form
 \begin{equation}
- (a_{UV}-a_{IR})\int d^d x\,\sqrt{-\hat g}\tau(x)\,{\hat E}_d.
\end{equation}

In deriving this result we have used the fact that
 \begin{equation}
a_{UV}-a_{IR} \propto \frac{1}{\kappa^2}\left(\Luv^{d-1}-\Lir^{d-1}\right),
\end{equation}
with $L_{UV}$ and $L_{IR}$  denoting the radii of the two asymptotically $AdS$ spaces and $\kappa$ is the $d$ dimensional Newton constant.
 Since according to \cite{Freedman:1999gp}   $\Luv\geq \Lir$,\footnote{For this to be true, the energy-momentum tensor of the bulk matter fields should satisfy the null energy condition $T_{MN}u^M u^N\geq 0$, with $g_{MN}u^M u^N=0$.} it  indeed  follows that the coefficient of the anomalous term $a_{UV}-a_{IR} \geq 0$ is positive.
 Note that we do not give here an independent holographic proof of this positivity since we use the null energy condition. However, it does show that we have identified consistently the holographic dilaton mode. The anomalous term in curved space implies the existence of a Wess-Zumino term in flat space for the dilaton with the same coefficient and a term $(\partial\tau)^d$. A consequence is that the scattering amplitude of $d$ dilatons should satisfy a positivity condition. This result is in full accordance with  the ``$a$-theorem" of \cite{Komargodski:2011vj}.
  We would like to emphasize that whereas in field theory such a statement has been proven in two and four dimensions, the holographic $a$-theorem holds at any even dimension.

In addition to the study of the holographic ``$a$-theorem", we analyze in detail the manifestations of the Goldstone theorem related to the spontaneous breaking of conformal symmetry by computing the one and two point functions of the  operator dual of the scalar  and the energy momentum tensor.
To implement this, we consider a suitable potential for the scalar field which admits the appropriate asymptotic behavior in the UV corresponding to a vacuum expectation value (vev) of a dual operator $\mathcal{O}$ of the $CFT_{UV}$. More precisely, the non-normalizable mode of the scalar field is set to zero, but the normalizable mode can have any value. There is a family of classical gravity solutions satisfying these boundary conditions and that at the same time are regular in the bulk. The near-horizon geometry is $AdS$, meaning that there is a conformal symmetry in the IR of the dual field theory. The renormalized on-shell action of the gravitational theory is independent of the value of the normalizable mode, hence there is a ``moduli space'' for the field theory dual with spontaneous symmetry breaking.
We first  expand   Einstein's equations to linear order in the fluctuations, fixing a particular (radial) gauge.
We then analyze the tensor and scalar  fluctuations at low momenta away from the horizon and compare with the finite momentum solution close to the horizon. We fix the boundary conditions at the horizon to be ingoing, so that we compute retarded correlators or, by analytic continuation, Euclidean correlators. By carefully  matching the  two asymptotic behaviors
we can determine the full solution to leading order in momentum. The spectrum of fluctuations reveals a zero mode corresponding to a change of the expectation value in agreement to what is expected when the symmetry is spontaneously broken. Evaluating the on-shell action on these solutions we can compute correlation functions using the standard holographic prescription. First, we confirm that the operator dual to the scalar field has a non-zero expectation value $\vev{\cO}$, while the trace of the energy-momentum tensor vanishes.
The two-point function of the energy-momentum tensor reads
\begin{equation}\label{TTcorrintro}
\vev{T^{\mu\nu}(-q) T^{\alpha\beta}(q)}\propto \frac{\Lir^d}{\kappa^2\Luv}\frac{1}{(q^2)^2}\Pi^{\mu\nu,\alpha\beta}(\sqrt{-q^2})^d\log(\Lir \sqrt{-q^2})^2,
\end{equation}
where $q^\mu$ is the four momentum and   $\Pi^{\mu\nu,\alpha\beta}$ is a kinematic factor depending on the momentum and the metric which is transverse and traceless. The explicit expression is given in \eqref{projector}.
The two-point function involving the energy-momentum tensor and the scalar operator is given by
\begin{equation}\label{TOcorrintro}
\left< T^{\mu\nu}(-q) \mathcal{O}(q) \right>  =\frac{\Delta_{UV}\left<\mathcal{O}\right>}{2(d-1)} \frac{1}{q^2} P_T^{\mu\nu}-\vev{\cO}\eta ^{\mu\nu}.
\end{equation}
where $P_T^{\mu\nu}$ is a kinematic factor transverse to the momentum (see (\ref{projector})).
This result  is in agreement  with the field theory  expectation based on Goldstone's theorem.
For the two-point function of the scalar  operator we find
\begin{equation}\label{OOcorrintro}
\vev{\cO(-q)\cO(q)} \propto \frac{\kappa^2\Lir^{d+2-2\Delta_{IR}}}{\Luv}\vev{\cO}^2(\sqrt{-q^2})^{d-2\Delta_{IR}}.
\end{equation}
This result exhibits a long range interaction, which is consistent with Goldstone's theorem (a more detailed explanation is given in Sec.~\ref{sec:compare}), but the zero-momentum singularity is not a pole, as one would expect to find for a free dilaton mode, and in fact the singularity leads to an IR divergence that is at odds with unitarity.
Recall that na\"\i vely one would expect to find in the IR a $CFT_{IR}~\times$ free dilaton. We comment more on this in the Discussion section.
The absence of a pole is in contrast with the probe brane analysis \cite{Elvang:2012st}, where the dilaton is a mode which essentially decoupled from the remaining massless degrees of freedom.

The paper is organized as follows. In Section \ref{sec:RG} the basic setup  of the holographic RG flows which is  analyzed  in this paper is established. We also explain which assumptions we make to simplify the analysis. Section \ref{sec:matching} is devoted to the manifestation of the conformal anomaly matching in holography. We first briefly review the picture of the boundary  field theory. A compensator field (spurion) is introduced to maintain conformal invariance of the field theory coupled to an external curved space-time. The anomalous terms of the generating functional are written down and the corresponding ``$a$-theorem" \cite{Komargodski:2011vj} is stated. Back to the holographic dual, we show that   the generating functional of the field theory  includes an anomaly term whose coefficient  is given by the difference between the $a$-anomaly of the UV and of the IR, in full accordance with  the ``$a$-theorem of \cite{Komargodski:2011vj}.
The holographic spontaneous breaking of conformal invariance is discussed in Section \ref{sec:spont}, where our strategy to find approximate solutions valid at small momentum is described.
  We determine the tensor and scalar  fluctuations
 at low momenta  and  extract the holographic correlators
 $\vev{\cO}, \vev{\cO\cO},\vev{T^{\mu\nu}\cO}$ and $\vev{T^{\mu\nu}T^{\alpha\beta}}$.
In Section \ref{sec:promote} we promote the spurion field  to a dynamical mode by introducing a new set of boundary conditions for the metric.  We conclude, summarize our assumptions and discuss several open questions in section \ref{sec:discuss}.
We end this paper with four appendices. In the first we elaborate on the equations of motions of the fluctuations. In appendix \ref{app:matching} we elaborate on the matching procedure and explicitly show the existence of an overlapping region between the boundary and the near-horizon region.  Appendix \ref{app:coulomb} is devoted to an application of our  matching procedure  to the case of the Coulomb branch of the ${\cal N}=4$ SYM theory  that was previously analyzed using a different procedure in \cite{Papadimitriou:2004rz,DeWolfe:2000xi,Arutyunov:2000rq,Mueck:2001cy}. In appendix \ref{app:toy} we present a family of toy models that describe an RG flow between two fixed points.

\section{Holographic RG flows}\label{sec:RG}
In most known examples, theories with a holographic dual have a classical weakly-coupled gravitational description when they are in a strong-coupling and large-$N$ limit, the canonical example being $\cN=4$ $SU(N)$ super Yang-Mills at large 't Hooft coupling. As we will see later we will study toy models of classical gravity coupled to matter with no known field theory description. We will assume that if a dual description of any of the models exists, the corresponding field theory dual will also be strongly coupled and in some sort of large-$N$ limit, as in the known examples.
A holographic dual of a theory with an RG flow between two fixed points will be a geometry that interpolates between two anti-de Sitter spaces with different radii. In Gaussian coordinates, the metric will be
\begin{align}
\notag &ds^2_{UV}=dr^2+e^{2r/\Luv}\eta_{\mu\nu}dx^\mu dx^\nu, \ \ r\to +\infty,\\
&ds^2_{IR}=dr^2+e^{2r/\Lir}\eta_{\mu\nu}dx^\mu dx^\nu, \ \ r\to-\infty,
\end{align}
where, imposing a null energy condition on the bulk matter fields the radius close to the boundary $r\to \infty$ is larger than the radius close to the horizon $r\to-\infty$: $\Luv>\Lir$. The radial coordinate $r$ maps to the RG scale of the dual theory, while the coordinates $x^\mu$ map to the space-time where the dual theory lives. Other RG flows where the IR geometry is not AdS are also possible, however, they do not describe a CFT in the far IR and for this reason we will no consider them.

Asymptotically the geometry has the following isometry
\begin{equation}
r\to r+\lambda, \ \ x^\mu\to e^{-\lambda/L}x^\mu,
\end{equation}
where $\lambda$ is a constant and $L=\Luv$ or $L=\Lir$ depending on which limit we take. A shift in the $r$ coordinate is then equivalent to a dilatation transformation of the space-time coordinates.

There are examples in supergravity of geometries that interpolate between two different AdS spaces (e.g. \cite{Freedman:1999gk,Freedman:1999gp,Girardello:1999bd,Bigazzi:2001ta,Berg:2001ty,Halmagyi:2005pn,Hotta:2008xt}), but they describe RG flows driven by a relevant deformation of the Lagrangian and not spontaneous breaking of conformal symmetry. Since we are interested mainly in the latter, we will construct our own holographic toy models with the desired properties, keeping the discussion as general as possible, so the results could easily be extended to particular examples in supergravity in case some are eventually found. Our construction is inspired by similar models that have been studied in the past   \cite{Martelli:2001tu,D'Hoker:2002aw,Freedman:2003ax}.

It is relatively easy to construct holographic models dual of a flow between fixed points, taking as inspiration some of the known supergravity flows. The simplest case is a scalar field with some potential $V$ coupled to Einstein gravity in $d+1$ dimensions.
\begin{equation}\label{EHaction}
S_{EH}=\frac{1}{\kappa^2}\int d^{d+1} x \sqrt{-g} \left( -\frac{1}{2}R-\frac{1}{2}\partial_M \phi \partial^M \phi -V(\phi)\right).
\end{equation}
Note that for this theory the null energy condition is automatically satisfied.

The requirements on the potential are
\begin{itemize}
\item[a)] There are at least two critical points where the potential is negative, one is a maximum and the other is a minimum.
\item[b)] There is a classical solution that interpolates between the two critical points.
\end{itemize}
The general case is still too complicated, but in certain cases it is possible to find a system of first order equations whose solutions are also solutions of the full second order system. That is the case when the potential is written in terms of a superpotential $W$\footnote{Locally in field space the potential can always be written in terms of a superpotential, by solving the equation \eqref{potential} understood as a differential equation. For each solution one has a different RG flow geometry, that could be dual to a theory with explicit or spontaneous breaking of conformal invariance \cite{Skenderis:1999mm,Skenderis:2006jq,Skenderis:2006rr}. Here we assume a prior knowledge of the superpotential, which is assumed to be valid in all configuration space, but the analysis could be extended to situations where this is not the case. We thank Ioannis Papadimitriou for pointing this out.}
\begin{equation}\label{potential}
V=\frac{1}{2}\left[(\partial W(\phi))^2-\frac{d}{d-1}W(\phi)^2 \right].
\end{equation}
One introduces an ansatz for the metric with Poincar\'e invariance along the space-time directions
\begin{equation}\label{gaussmet}
ds^2=dr^2+e^{2A}\eta_{\mu\nu}dx^\mu dx^\nu,
\end{equation}
and the first order equations that give solutions for the Einstein plus scalar equations of motion are
\begin{equation}\label{floweq}
\phi'=-\partial W, \ \ A'=\frac{W}{d-1},
\end{equation}
where primes denote radial derivatives. Critical points of the superpotential are also critical points of the potential, so in order to have a flow we should demand that the superpotential has two critical points. Without loss of generality we can assume that one of the critical points is at $\phi=0$,
\begin{equation}\label{superpot}
W\simeq \frac{d-1}{\Luv}+\frac{\Delta_{UV}}{2\Luv} \phi^2, \ \ V\simeq-\frac{d(d-1)}{2\Luv^2}+\frac{\Delta_{UV}(\Delta_{UV}-d)}{2\Luv^2}\phi^2.
\end{equation}
Here $\Delta_{UV}$ is the dimension of the operator dual to the field $\phi$. We will assume in the following that the dual operator is relevant, so $\Delta_{UV}<d$. In this case the potential has a maximum.

The second critical point will be at some value $\phi_m$, such that
\begin{equation}\label{superpot2}
W\simeq \frac{d-1}{\Lir}+\frac{d-\Delta_{IR}}{2\Lir} (\phi-\phi_m)^2, \ \ V\simeq-\frac{d(d-1)}{2\Lir^2}+\frac{\Delta_{IR}(\Delta_{IR}-d)}{2\Lir^2}(\phi-\phi_m)^2.
\end{equation}
We will demand that $\Delta_{IR}>d$, so the dual operator becomes irrelevant in the IR fixed point and the potential has a minimum. Solutions can be simply obtained by using the field $\phi$ as radial coordinate,
\begin{equation}
ds^2=\frac{d\phi^2}{(\partial W)^2}+e^{2A(\phi)}\eta_{\mu\nu} dx^\mu dx^\nu, \ \ \partial A(\phi)=-\frac{1}{d-1}\frac{W}{\partial W}.
\end{equation}
A family of simple superpotentials satisfying the requirements is given in Appendix \ref{app:toy}. In Figure \ref{figpot} we have plotted an example of a superpotential with the required properties and its corresponding potential.

Note that the asymptotic value of the scalar field is proportional to the expectation value of the scalar operator, and that one would recover pure AdS if this value is set to zero. An explicit way to see this is first expanding the superpotential when $\phi\to 0$ and then solving the equation of motion for $\phi$,
\begin{equation}
\phi\simeq C_1 e^{-\Delta_{UV} r/L_{UV}}+\cdots,
\end{equation}
where $C_1 \propto \left\langle {\cal O} \right\rangle$.  Higher order terms can be found continuing the expansion of the superpotential and solving iteratively the equation, they have higher powers of $C_1 e^{-\Delta r/L}$.

With this solution we can solve for $A$
\begin{equation}
A=A_0+\frac{r}{L_{UV}}-\frac{1}{4(d-1)}C_1^2 e^{-2\Delta_{UV} r/L_{UV}}+\cdots.
\end{equation}
Therefore, in the limit where the expectation value vanishes, $C_1\to 0$, the geometry becomes
\begin{equation}
\phi \to 0, \ \ \ A\to A_0+\frac{r}{L_{UV}}.
\end{equation}
Which is just pure AdS with no scalar turned on. A technical but important point is that we will take the dimensions $\Delta_{UV}$ and $\Delta_{IR}$ not to be integers, the
reason being that the analysis simplifies and one avoids potential contributions to the Weyl anomaly from multitrace operators. We will also impose that the number of dimensions is even (since there is no anomaly in odd dimensions) and that $d>2$ for the calculation of correlators (since for $d=2$ one does not expect to have a dilaton \cite{Coleman:1973ci}).

\FIGURE[t]{
\includegraphics[width=6cm]{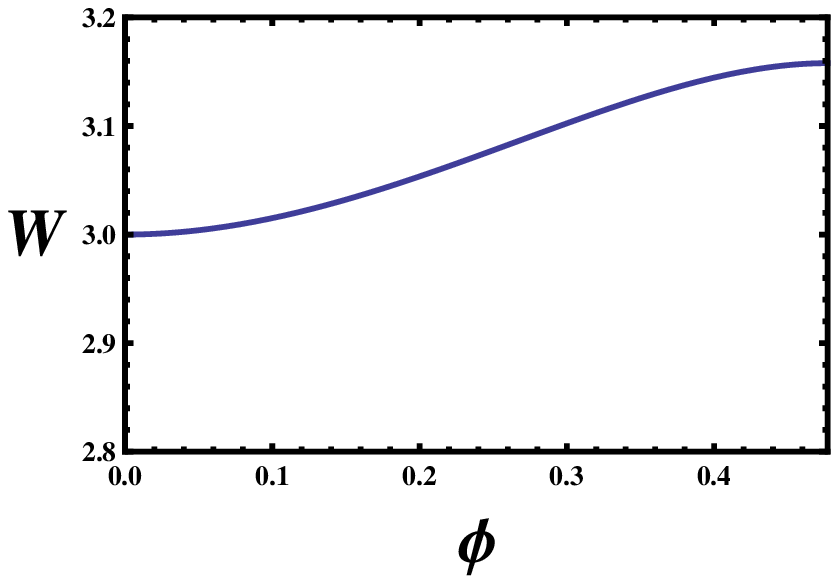}
\includegraphics[width=6cm]{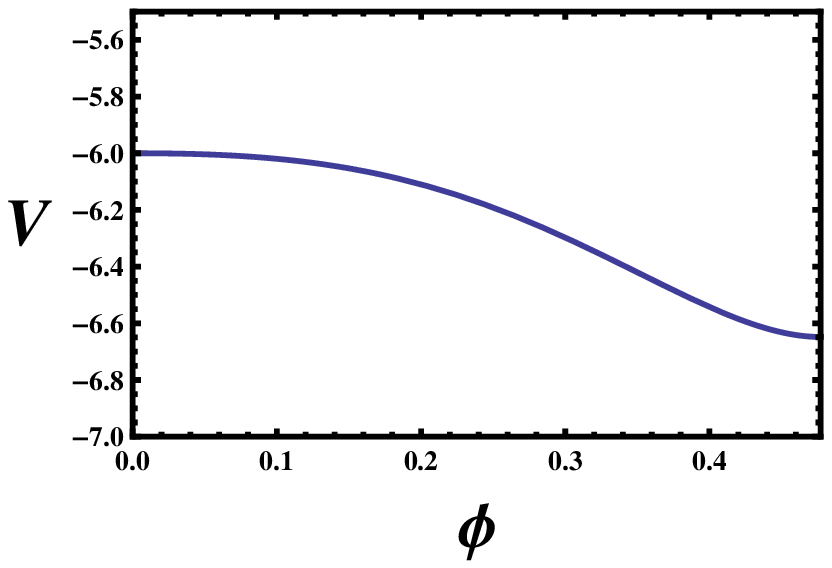}
	\caption{ An holographic RG flow interpolates between the minimum and the maximum of the superpotential on the left side, or equivalently between the maximum and the minimum of the potential on the right side.
	}
	\label{figpot}
}

The construction can be generalized to several scalars, that may have a non-trivial metric in configuration space. The basics will be the same, the flow between two fixed points will be dual to a solution to the equations interpolating between two critical points of the superpotential with the right properties. In this case one has to worry about the directions transverse to the flow in configuration space, and make sure that there are no unstable modes.

\section{Conformal anomaly matching in holography}\label{sec:matching}

Our purpose in this section is to show explicitly how the anomaly matching arguments presented in \cite{Komargodski:2011vj,Komargodski:2011xv}\footnote{See also \cite{Schwimmer:2010za} for the original arguments regarding anomaly matching in theories with spontaneously broken conformal invariance.} for the generating functional are realized in gauge/gravity models.

\subsection{Field theory preliminaries}

Consider first a conformally invariant theory. In the presence of a background metric conformal invariance is explicitly broken. However, one can introduce a compensator field such that the theory is invariant under an extended symmetry. The compensator field, or spurion, will enter as a Weyl factor of the background metric, as was discussed in \cite{Komargodski:2011vj,Komargodski:2011xv},
\begin{equation}
g_{\mu\nu} \to \hat g_{\mu\nu}=e^{2\tau} g_{\mu\nu}.
\end{equation}
The extended symmetry involves both a Weyl rescaling of the metric and a shift of the spurion:
\begin{equation}\label{spuriont}
g_{\mu\nu}\to e^{2\sigma}g_{\mu\nu}, \ \ \tau\to \tau-\sigma.
\end{equation}
Correlation functions of the energy-momentum tensor in the CFT can be obtained by deriving a generating functional (generally non-local), $\Gamma$, with respect to the background metric. The generating functional should have the same symmetries as the original theory, so it should be a diffeomorphism-invariant functional of the dressed metric $\hat g_{\mu\nu}$. In the absence of anomalies this will be the end of the story. However, when the number of dimensions $d$ is even, a CFT in curved space has in general a conformal anomaly. The generating functional should reproduce the anomaly, which implies that it is not completely invariant under the transformation \eqref{spuriont} but it transforms as
\begin{equation}
\left.\delta_\sigma  \Gamma\right|_{\tau=0} = \int d^d x\sqrt{-g}  \sigma {\cal A}_d,
\end{equation}
where ${\cal A}_d$ is a polynomial of the curvature and its derivatives with $d$ derivatives of the metric. The anomaly in the generating functional coincides with the expectation value of the trace of the energy-momentum tensor.
\begin{equation}
\vev{T^\mu_{\ \mu}}=\left.g_{\mu\nu}\frac{2}{\sqrt{-g}}\frac{\delta \Gamma}{\delta g_{\mu\nu}}\right|_{\tau=0}={\cal A}_d.
\end{equation}
For instance, in  two-dimensions the anomaly polynomial is
\begin{equation}\label{Atwo}
{\cal A}_2=-\frac{c}{24\pi}R,
\end{equation}
where $c$ is the central charge of the CFT. In four and larger dimensions the number of independent coefficients is larger. Concretely, in four dimensions there are three possible independent contributions to the anomaly
\begin{equation}\label{Afour2}
{\cal A}_4=c \cW^2-a E_4+b \square R,
\end{equation}
where $\cW^2$ is the Weyl tensor squared and $E_4$ is the Euler density. The last term is not directly related to an anomaly, but it is a contact term that may appear in the energy-momentum tensor correlators and that can be eliminated through the addition of a local counterterm. The first and second terms are labeled by the coefficients $c$ and $a$. The $a$-anomaly (known as type A) is the equivalent to the conformal anomaly in two dimensions, and it shares some properties with chiral anomalies: it is determined by a one-loop diagram with $d/2+1$ legs, there is no ultraviolet divergence involved and there is a set of descent equations related to the anomaly \cite{Boulanger:2007ab}. The $c$-anomaly (known as type B) on the other hand is related to UV divergences of correlation functions (see e.g. \cite{Schwimmer:2010za}). In dimensions larger than four there is also a type A anomaly proportional to the Euler density. Type B anomalies also appear but their number increases with the number of dimensions, their classification is based on the construction of Weyl-invariant polynomials of the curvature, although in the presence of other sources there can be additional B anomalies.

Now consider a situation where conformal symmetry has been broken, either spontaneously or explicitly, and the theory flows to an IR fixed point. In the latter case it is still possible to maintain a symmetry of the form \eqref{spuriont}, by dressing dimensionful quantities with the spurion field. In other words, any mass scale $m$ should be multiplied by a factor $e^{-\tau}$. In this way the generating functional in both cases takes essentially the same form. The generating functional will suffer in general from IR divergences due to the massless degrees of freedom of the IR CFT. One can avoid this by introducing an IR regulator $\mu$ and integrating out all the degrees of freedom above this scale, but keeping the degrees of freedom of the IR CFT. The resulting partition function will take then the form
\begin{equation}
Z=e^{\Gamma[\mu; \hat g_{\mu\nu}]}Z_{IR\; CFT}[\hat g_{\mu\nu}].
\end{equation}
With the IR regulator, $\Gamma[\mu; \hat g_{\mu\nu}]$ is a local functional of the dressed metric and couplings.

From the anomaly matching arguments of \cite{Schwimmer:2010za} and \cite{Komargodski:2011vj,Komargodski:2011xv}, the variation of the total partition function is determined by the coefficients of the conformal anomaly in the UV theory, $a_{UV}$ and $c_{UV}$. However, the IR CFT in general contributes to the anomaly with different coefficients, $a_{IR}$ and $c_{IR}$. Therefore, the generating functional $\Gamma[\mu; \hat g_{\mu\nu}]$ must account for the difference. This implies that there is a term in the generating functional whose variation gives
\begin{equation}
\delta_\sigma \Gamma[\mu; \hat g_{\mu\nu}] = \int d^d x \sqrt{-g} \sigma \left(\cA_d^{UV}-\cA_d^{IR}\right).
\end{equation}
For instance, in two dimensions
\begin{equation}
\Gamma[\mu; \hat g_{\mu\nu}] \supset -\frac{c_{UV}-c_{IR}}{24\pi}\int d^2 x \sqrt{- g} \tau\left( R+(\partial\tau)^2\right).
\end{equation}
The terms in four and six dimensions can be found in \cite{Komargodski:2011vj} and \cite{Elvang:2012st} respectively. In general, the coefficients of the anomalous terms coming from the variation of the functional should be proportional to the difference between the coefficients of the anomalies in the UV and IR CFTs. In two and four dimensions this allows to prove a $c$- or $a$- theorem \cite{Komargodski:2011vj,Komargodski:2011xv},
\begin{equation}
c_{UV}\geq c_{IR}, \ \ a_{UV}\geq a_{IR}.
\end{equation}
Of course in two dimensions the $c$-theorem was proven long ago by Zamolodchikov using a different approach \cite{Zamolodchikov:1986gt}. The new proof is based on the analytic properties of scattering amplitudes, unfortunately it seems that the extension of these arguments to six dimensions are not straightforward and a similar statement does not exist there \cite{Elvang:2012st}.

This is in contrast with holographic models, where there are various existing proofs of $a$-theorems in arbitrary number dimensions \cite{Alvarez:1998wr,Freedman:1999gp,Myers:2010tj}. We will show that the proof given in \cite{Freedman:1999gp} for RG flow geometries implies that the coefficient of the anomalous term in the generating functional is positive. Therefore, there may still be an argument purely within field theory that fixes the sign of the coefficient in six dimensions, even though it would be more subtle than in two and four dimensions.

\subsection{Spurion field for Weyl transformations}

We will now explain how to introduce a spurion field in a holographic dual. We start with the RG flow metric \eqref{gaussmet} and do the change of coordinates $\rho=e^{-2A}$
\begin{equation}\label{FGmetric}
ds^2=\ell^2(\rho)\frac{d\rho^2}{4\rho^2}+\frac{1}{\rho}\eta_{\mu\nu}dx^\mu dx^\nu,
\end{equation}
where
\begin{equation}
\ell(\rho)=\frac{1}{A'},
\end{equation}
asymptotes $\Luv$ in the $\rho\to 0$ limit and $\Lir$ in the $\rho\to \infty$ limit. In these two limits the metric just becomes $AdS$ in Fefferman-Graham coordinates. We can generalize this form of the metric to arbitrary solutions of the equations of motion
\begin{equation}
ds^2=\ell^2(\rho,x)\frac{d\rho^2}{4\rho^2}+\frac{1}{\rho}g_{\mu\nu}(x,\rho)dx^\mu dx^\nu,\ \ \ \phi=\phi(x,\rho),
\end{equation}
where, close to the boundary $\rho\to 0$ the lapse factor becomes a constant $\ell\to \Luv$. In this limit the metric and the scalar field will have an expansion in powers of the radial coordinate (for $d$ even dimensions) \cite{deHaro:2000xn}
\begin{align}\label{metexp}
\notag g_{\mu\nu}(x,\rho)&=g_{\mu\nu}^{(0)}(x)+g_{\mu\nu}^{(2)}(x)\rho+\cdots+g_{\mu\nu}^{(d)}(x)\rho^{d/2} +h_{\mu\nu}^{(d)}(x)\rho^{d/2}\log\rho\\ \notag
&+\cdots+g_{\mu\nu}^{(d-\Delta_{UV})}(x)\rho^{d-\Delta_{UV}}+\cdots \\
\phi(x,\rho)&=\phi^{(d-\Delta_{UV})}(x)\rho^{(d-\Delta_{UV})/2}+\cdots+\phi^{(\Delta_{UV})}(x)\rho^{\Delta_{UV}/2}+\cdots
\end{align}
For odd $d$ dimensions, the logarithmic term proportional to $h_{\mu\nu}^{(d)}(x)$ is not present. As we will see this term is crucially related to the conformal anomaly.
We are assuming that $\Delta_{UV}$ is not an integer, otherwise there can be additional logarithmic terms in the expansion of the scalar field that will make the analysis more complicated.

In this expansion all the coefficients are completely determined by the leading terms $g_{\mu\nu}^{(0)}$ and $\phi^{(d-\Delta_{UV})}(x)$, up to the terms $g_{\mu\nu}^{(d)}(x)$ and $\phi^{(\Delta_{UV})}(x)$ for which the boundary conditions when $\rho\to\infty$ are also needed. The leading term in the expansion of the scalar field $\phi^{(d-\Delta_{UV})}(x)\neq 0$  maps to a source for the dual operator in the field theory and breaks explicitly conformal invariance. The scalar solution still vanishes at the boundary, so the asymptotic AdS form of the metric is not affected. From the perspective of the holographic dual we are introducing a relevant deformation, so the UV physics will not be affected by it.

The coefficients $g_{\mu\nu}^{(n)}$ have been computed in pure gravity and in theories with scalar fields as well \cite{deHaro:2000xn,Bianchi:2001de,Bianchi:2001kw,Papadimitriou:2004rz}.
Each term has $n$ derivatives, so the expansion above can also be seen as a derivative expansion. In particular, if the derivatives are small enough, it can be extended to values of $\rho$ that are not asymptotically close to $\rho=0$. This implies that for small enough derivatives
\begin{equation}
|\partial_\alpha g_{\mu\nu}^{(0)}|, \ \ |\partial_\alpha\phi^{(0)}| \ \ \ll \frac{1}{\rho_{IR}^{1/2}},
\end{equation}
the expansion above will also be valid in a region of the geometry where the space asymptotes $AdS$ with radius $\Lir$, that we will call the near-horizon region, even though a solution with explicit dependence on the space-time coordinate will most likely continue to a different geometry when $\rho\to \infty$.

Note that the solution in the near-horizon region is determined by the same leading term of the metric $g_{\mu\nu}^{(0)}$, but that the scalar field will have a different expansion
\begin{equation}
\phi(x,\rho)=\phi^{(d-\Delta_{IR})}(x)\rho^{(d-\Delta_{IR})/2}+\cdots+\phi^{(\Delta_{IR})}(x)\rho^{\Delta_{IR}/2}+\cdots
\end{equation}
From the perspective of the near-horizon AdS, $\phi^{(d-\Delta_{IR})}$ acts as a ``source'' although of course it is not independent of the boundary value $\phi^{(d-\Delta_{UV})}$, and the same is true for the subleading terms. For the explicit solution of the holographic RG flow, one sees that the scalar solution interpolates between the two scalings in a way determined by the superpotential.

In the case where the geometry is the dual of a theory with spontaneous breaking of symmetry, the source term will vanish $\phi^{(d-\Delta_{UV})}=0$. This is the case we have considered in the previous section, here we will keep the source term in order to make the discussion more general.

We now proceed to introduce the spurion field. Note that in field theory the spurion was introduced by making a Weyl transformation of the metric and of dimensionful quantities. The question is what is the realization of this in the gravity dual. The answer for infinitesimal Weyl transformations in pure AdS is the so-called Penrose-Brown-Henneaux diffeomorphism \cite{Penrose:1986ca,Brown:1986nw} introduced a holographic context in \cite{Imbimbo:1999bj}. We will generalize it to the RG flow geometries we introduced in the previous section and to non-infinitesimal transformations. The PBH transformation has also been used recently to construct the effective action of the dilaton from the action of probe branes embedded in $AdS$ \cite{Elvang:2012st}. The difference with this approach is that the probe brane represents a breaking of conformal invariance that is suppressed in the large-$N$ limit, so that to leading order the geometry that describes the UV fixed point is not affected and a decoupled dilaton shows up in the spectrum as a fluctuation of the brane. In the cases we study the breaking is not suppressed and the effect of the dilaton is seen indirectly in correlation functions.

In pure AdS space the metric is \eqref{FGmetric} with a constant $\ell^2(\rho)=L^2$. The infinitesimal PBH transformation takes the form
\begin{align}
&\rho\to e^{-2\tau}\rho\simeq (1-2\tau)\rho, \ \ \ x^\mu\to x^\mu+a^\mu,\\
&a^\mu=\frac{L^2}{2}\int_0^\rho d\rho' g^{\mu\nu}(x,\rho')\partial_\nu\tau(x),
\end{align}
and it maps a solution with boundary metric $g_{\mu\nu}^{(0)}(x)$ to a new solution where the original metric has changed by an infinitesimal Weyl transformation
\begin{equation}\label{varmet}
\delta g_{\mu\nu}^{(0)}(x)=+2\tau(x) g_{\mu\nu}^{(0)}(x).
\end{equation}
A similar infinitesimal transformation exists for the more general case \eqref{FGmetric}
\begin{align}\label{inftPBH}
&\rho\to e^{-2\tau}\rho\simeq (1-2\tau)\rho, \ \ \ x^\mu\to x^\mu+a^\mu,\\
&a^\mu=\frac{1}{2}\int_0^\rho d\rho' \ell^2(\rho')g^{\mu\nu}(x,\rho')\partial_\nu\tau(x),
\end{align}
The effect of this transformation over the leading term of the metric is again \eqref{varmet}. The leading terms of the scalar field transforms as
\begin{equation}
\delta \phi^{(d-\Delta_{UV})}=-(d-\Delta_{UV})\tau(x) \phi^{(d-\Delta_{UV})}, \ \ \  \delta \phi^{(d-\Delta_{IR})}=-(d-\Delta_{IR})\tau(x) \phi^{(d-\Delta_{IR})}.
\end{equation}
Subleading terms in the expansion of the metric and the scalar field \eqref{metexp} have a covariant dependence on the leading terms, so the effect of this transformation is to make them dependent on the combinations $g_{\mu\nu}^{(0)}+\delta g_{\mu\nu}^{(0)}(x)$, $\phi^{(d-\Delta_{UV})}+\delta \phi^{(d-\Delta_{UV})}$. There is however one term that does not change covariantly, due to the logarithmic term in the expansion of the metric \eqref{metexp}. We have then, that when $\rho\to 0$,
\begin{equation}
\delta g_{\mu\nu}^{(d)}=\frac{\delta g_{\mu\nu}^{(d)}}{\delta g_{\alpha\beta}^{(0)}}\delta g_{\alpha\beta}^{(0)}+\frac{\delta g_{\mu\nu}^{(d)}}{\delta\phi^{(d-\Delta_{UV})}}\delta \phi^{(d-\Delta_{UV})}-2\tau(x)h_{\mu\nu}^{(d)},
\end{equation}
while, when $\rho\to\infty$,
\begin{equation}
\delta g_{\mu\nu}^{(d)}=\frac{\delta g_{\mu\nu}^{(d)}}{\delta g_{\alpha\beta}^{(0)}}\delta g_{\alpha\beta}^{(0)}+\frac{\delta g_{\mu\nu}^{(d)}}{\delta\phi^{(d-\Delta_{IR})}}\delta \phi^{(d-\Delta_{IR})}-2\tau(x)h_{\mu\nu}^{(d)}.
\end{equation}

Introducing a spurion field requires to promote the infinitesimal PBH transformation to a finite transformation, including non-linear effects
\begin{align}
\rho\to e^{-2\tau}\rho+\delta \rho,\ \ x^\mu\to x^\mu+a^\mu,
\end{align}
where
\begin{equation}
a^\mu=\frac{1}{2}\int_0^{\rho e^{-2\tau}} d\rho' \ell^2(\rho')g^{\mu\nu}(x,\rho')\partial_\nu\tau(x)+\delta a^\mu.
\end{equation}
The boundary values of the fields are affected by this transformation, so in general the resulting solution will not interpolate between two AdS spaces any more. However, if the derivatives of $\tau(x)$ are small enough we can apply the same derivative expansion \eqref{metexp} in the near-horizon region up to some value of the radial coordinate $\rhir$.

Indeed, the non-linear generalization of the PBH transformation is consistent with an expansion of $\delta \rho$ and $\delta a^\mu$ in terms of the number of derivatives of $\tau$
\begin{align}
\notag &\delta \rho=\delta\rho^{(2)}(\partial^2\tau)+\delta\rho^{(4)}(\partial^4\tau)+\delta\rho^{(6)}(\partial^6\tau)+\cdots,\\
&\delta a^\mu=\delta a^\mu_{(3)}(\partial^3\tau)+\delta a^\mu_{(5)}(\partial^5\tau)+\delta a^\mu_{(7)}(\partial^7\tau)+\cdots
\end{align}
Each term is fixed by demanding that the metric keeps the form
\begin{equation}\label{FGmetric2}
ds^2=\ell^2(\rho e^{-2\tau})\frac{d\rho^2}{4\rho^2}+\frac{1}{\rho}e^{2\tau}\tilde g_{\mu\nu}(\rho e^{-2\tau},x)dx^\mu dx^\nu.
\end{equation}
For instance, the contribution to the shift in $\rho$ that is second order in derivatives is determined by the equation
\begin{equation}
\left[e^{2\tau}\partial_\rho+\frac{\ell'}{\ell} -\frac{1}{\rho e^{-2\tau}}\right]\delta \rho^{(2)} =-\frac{\rho e^{-2\tau}}{2}g^{\alpha\beta}\partial_\alpha\tau\partial_\beta\tau.
\end{equation}
Although straightforward, this expansion can be quite cumbersome. It is also not completely unambiguous, for instance the equation above has a homogeneous solution
\begin{equation}
\delta \rho^{(2)}_H= \rho e^{-2\tau}\frac{1}{\ell} f(x),
\end{equation}
for some arbitrary function $f(x)$ that has to be second order in derivatives in order to be consistent with the expansion, an example is $f(x)=e^{-2\tau}g_{(0)}^{\alpha\beta}\partial_\alpha\tau\partial_\beta\tau$ . Close to the boundary, this can be seen as a field redefinition of $\tau$ adding higher derivative terms
\begin{equation}
\tau\to \tau+\frac{1}{\Luv} e^{-2\tau}g_{(0)}^{\alpha\beta}\partial_\alpha\tau\partial_\beta\tau+O(\partial^4).
\end{equation}

The final effect of the transformation is to do a finite Weyl rescaling of the sources for the metric and the scalar field, both at the boundary and in the near-horizon region
\begin{align}\label{dressedsources}
\notag & g_{\mu\nu}^{(0)}(x)\to e^{2\tau(x)} g_{\mu\nu}^{(0)}(x)\equiv \hat g_{\mu\nu},\\
\notag & \phi^{(d-\Delta_{UV})}\to e^{-(d-\Delta_{UV})\tau(x)} \phi^{(d-\Delta_{UV})}\equiv \hat \phi_{UV},\\ & \phi^{(d-\Delta_{IR})}\to e^{-(d-\Delta_{IR})\tau(x)} \phi^{(d-\Delta_{IR})}\equiv \hat \phi_{IR}.
\end{align}
All other terms in the derivative expansion depend on these sources and transform covariantly according to the Weyl transformation, except for the anomalous contribution
\begin{equation}\label{shiftmetric}
g_{\mu\nu}^{(d)}[g_{\mu\nu}^{(0)},\phi^{(d-\Delta)}]\to g_{\mu\nu}^{(d)}[\hat g_{\mu\nu},\hat \phi]-2\tau(x)h_{\mu\nu}^{(d)}[\hat g_{\mu\nu},\hat \phi]
\end{equation}
The additional contribution appears only in even $d$ dimensions, and is due to the logarithmic term in the expansion of the metric in \eqref{metexp}.

\subsection{Anomalous contribution to the generating functional}

Having established how the spurion should be introduced in the holographic dual, we can now derive the anomalous contribution to the generating functional and check if the results agree with the expectation from field theory.

According to the AdS/CFT correspondence, the generating functional of the field theory is the on-shell action of the gravitational theory, as a function of the boundary values of the fields. We will assume a large-$N$, strong coupling approximation, such that the on-shell action reduces to the classical gravitational action \eqref{EHaction} evaluated on the solutions. We will define the on-shell action with cutoffs in the radial direction, $\rhuv\ll 1$ and $\rhir\gg 1$
\begin{align}\label{action}
\notag S[\rhuv,\rhir]&=-\frac{1}{2\kappa^2}\int d^d x\int_\rhuv^\rhir d\rho \sqrt{-G} \left(R+\partial_M \phi \partial^M \phi +2V(\phi)\right)\\ &+\frac{1}{\kappa^2}\left.\int d^{d} x \sqrt{-\tilde G} K\right|_{\rhir}-\frac{1}{\kappa^2}\left.\int d^{d} x \sqrt{-\tilde G} K\right|_{\rhuv}.
\end{align}
On the second line we have added Gibbons-Hawking terms, since we will do variations of the classical solution that affect the value of the metric at both cutoffs. Here $K$ is the extrinsic curvature, defined from the metric components $G_{\mu\nu}=\frac{1}{\rho}g_{\mu\nu}$. $\sqrt{-\tilde G}$ is the square root of the determinant of $G_{\mu\nu}$, while $\sqrt{-G}$ is the square root of the determinant of the full metric. $N^2=g_{\rho\rho}=\frac{\ell^2}{4\rho^2}$ is the lapse function. The explicit form of the extrinsic curvature is
\begin{equation}
K_{\mu\nu}=\frac{1}{2N} \partial_\rho G_{\mu\nu}=\frac{1}{\ell}\left(\partial_\rho-\frac{1}{\rho}\right) g_{\mu\nu}, \ \
K=G^{\mu\nu} K_{\mu\nu}=\rho g^{\mu\nu}K_{\mu\nu}.
\end{equation}

In anti-de Sitter space, on-shell action has the form \cite{Henningson:1998gx,Imbimbo:1999bj}
\begin{equation}\label{intS}
S=-\frac{L}{2\kappa^2}\int_{\rho_{UV}} d\rho d^d x\,\rho^{-\frac{d}{2}-1}\sqrt{-g^{(0)}}\,b(x,\rho),
\end{equation}
where, close to the boundary $\rho\sim \rho_{UV}\to 0$,
\begin{equation}
b(x,\rho)=\sum_{n=0}^\infty b_{2n}(x)\rho^n+\cdots.
\end{equation}
Where the dots refer to possible contributions that are not integer powers of $\rho$, which can appear in the presence of matter fields in the bulk. From this expansion one can compute the form of the divergent contributions to the on-shell action. Note that in even dimensions there is a divergent logarithmic contribution from the order $d$ term,
\begin{equation}
S_{\rm div}\supset \frac{1}{\kappa^2}\int d^d x\, \log(\rho_{UV}) b_{d}(x),
\end{equation}
that in fact is proportional to the anomaly
\begin{equation}\label{bd}
b_d=-\frac{d}{2}L^{d-2} \cA_d.
\end{equation}
Where here and in the following the anomaly polynomial $\cA_d$ has been defined in such a way that coefficient of the Euler density $E_d$ is one. The divergence is eliminated by adding a counterterm $S_{ct}=-S_{\rm div}$.

Under an infinitesimal PBH transformation the $\rho$ coordinate is rescaled by a $e^{-2\tau}$ factor, so that a change of variables in the integral appearing in \eqref{intS} puts this factor in the limits of integration $\rhuv\to \rhuv e^{-2\tau}$, $\rhir\to \rhir e^{-2\tau}$. Then, the on-shell action is shifted by a finite term proportional to the anomaly
\begin{equation}
\delta S_{\rm div} =-\frac{dL^{d-1}}{\kappa^2}\int d^d x\,\tau(x)\sqrt{-g^{(0)}} \cA_d.
\end{equation}
This can be compensated by a change of the finite part of the action if one combines the PBH transformation with a Weyl transformation of the boundary metric, in such a way that the boundary metric is left invariant. Note that this term is scheme-dependent, in a different scheme one could choose the cutoffs to depend on the spacetime coordinates, $\rhuv\to \rhuv e^{-2\sigma}$ and $\rhir\to \rhir e^{-2\sigma}$,
in such a way that the value of $\tau$ is shifted $\tau\to \tau+\sigma$. Demanding that there are no anomalous contributions before performing the PBH transformation will fix the scheme to $\sigma=0$.

In our case the action takes the form
\begin{equation}
S=-\frac{1}{\kappa^2}\int_{\rho_{UV}}^{\rho_{IR}} d\rho d^d x\,\ell(\rho)\rho^{-\frac{d}{2}-1}\sqrt{-g^{(0)}}\,b(x,\rho).
\end{equation}
We can expand the function $b(x,\rho)$ in the $AdS$ regions
\begin{align}
&\rho\sim \rhuv, \ \ b(x,\rho)=\sum_{n=0}^\infty b_{2n}^{UV}(x)\rho^n,\\
&\rho\sim \rhir, \ \ b(x,\rho)=\sum_{n=0}^\infty b_{2n}^{IR}(x)\rho^n.
\end{align}
The UV expansion is the usual $\rho\to 0$ expansion. The IR expansion is a derivative expansion that is valid for finite $\rho$ deep in the IR $AdS$ region but not all the way to the horizon, where the expansion breaks down. The coefficients $b_d^{UV}(x)$ and $b_d^{IR}(x)$ take the same form as \eqref{bd}, with $L$ changed to $\Luv$ and $\Lir$, respectively. The factor $\ell(\rho)$ becomes a constant, $\Luv$ or $\Lir$, plus $\rho$-dependent corrections. Since the flow is driven by the matter fields, these contributions take the form of non-integer powers of $\rho$.

From these expansions we can infer that there are contributions proportional to the anomaly coming from both the UV and IR cutoff.
\begin{equation}
S\supset -\frac{d}{2\kappa^2}\int d^d x\, \sqrt{-g^{(0)}}\left(\Lir^{d-1}\log(\rhir)-\Luv^{d-1}\log(\rhuv)\right)\cA_d.
\end{equation}
Following the same arguments as for the pure $AdS$ case, a PBH transformation will add to the action a term linear in $\tau$ and proportional to the difference between the UV and IR anomalies
\begin{equation}\label{anomterm1}
\delta S=-\frac{d}{\kappa^2}\left(\Luv^{d-1}-\Lir^{d-1}\right)\int d^d x\,\tau(x) \sqrt{-\hat g^{(0)}}\,\hat \cA_d.
\end{equation}
The contribution to the energy-momentum tensor of the theory can be computed from a variation of the expression above with respect to the metric. We will now give an alternative derivation that will serve as a cross-check of our result. Using the equations of motion, the on-shell action can also be expressed as
\begin{equation}\label{varonshell}
S[\rhuv,\rhir]=-\frac{1}{\kappa^2}\int d^d x\int_\rhir^\rhuv d\rho \sqrt{-G} \left(\frac{1}{2}\pi^{\mu\nu} \partial_\rho G_{\mu\nu}+\pi_\phi \partial_\rho \phi\right).
\end{equation}
This has the usual form of a classical action with a vanishing Hamiltonian, a well-known result in the ADM formalism that has also been used in the Hamilton-Jacobi formulation of RG flows \cite{deBoer:1999xf,Papadimitriou:2004rz}. The conjugate momenta to the metric and the scalar field are defined as
\begin{equation}
\pi^{\mu\nu}=\frac{\kappa^2}{\sqrt{-G}}\frac{\delta S}{\delta \partial_\rho G_{\mu\nu}}= K^{\mu\nu}-KG^{\mu\nu}, \ \ \ \pi_\phi=\frac{\kappa^2}{\sqrt{-G}}\frac{\delta S}{\delta \partial_\rho \phi}=\partial_\rho \phi.
\end{equation}
Because the action is on-shell, a variation of the action only has contributions from the cutoffs
\begin{equation}\label{variation}
\delta S[\rhuv,\rhir]=-\left.\frac{1}{\kappa^2}\int d^d x\,\sqrt{-G} \left(\frac{1}{2}\pi^{\mu\nu} \delta G_{\mu\nu}+\pi_\phi \delta \phi\right)\right|^{\rhuv}_{\rhir}.
\end{equation}
The variation at the boundary should vanish for dynamical modes, this corresponds to the Dirichlet boundary condition $\delta \phi=0$, $\delta G_{\mu\nu}=0$.

The regulated expectation value of the energy-momentum tensor is
\begin{equation}
T_{\mu\nu}^{\rm reg}=-\frac{1}{\kappa^2}\rho^{-\frac{(d-2)}{2}}\pi_{\mu\nu}.
\end{equation}
At the boundary $\rhuv\to 0$, this is divergent and one needs to add counterterms that cancel those divergences, which can be done by adding covariant local terms at the cutoff \cite{deHaro:2000xn}. In the near-horizon region the regulated energy-momentum tensor is finite and no additional contributions should be added.

After eliminating the divergences, the $\rhuv\to 0$ limit leaves only finite contributions to the energy-momentum tensor. After the holographic renormalization procedure has been implemented, the variation of the action takes the form
\begin{equation}
\delta S = \int d^d x \sqrt{-g^{(0)}}\left[\frac{1}{2}\vev{T^{\mu\nu}}\delta g^{(0)}_{\mu\nu}+\vev{\cO}\delta \phi^{(d-\Delta_{UV})} \right].
\end{equation}
In our case the expectation value of the energy-momentum tensor has contributions from the boundary and the near-horizon region. Near-horizon contributions depend in general on $\rhir$, that acts as an IR regulator. The exception is the order $d$ term (with no logarithmic factors), that has the same form as the renormalized boundary contribution. Using the results of \cite{Henningson:1998gx, deHaro:2000xn}, the contributions of order $d$ are
\begin{equation}
\vev{T_{\mu\nu}}_d[g^{(0)},\phi^{(d-\Delta_{UV})}] = \frac{d}{2\kappa^2}\left(\Luv^{d-1}-\Lir^{d-1}\right)e^{-(d-2)\tau}g^{(d)}_{\mu\nu}[g^{(0)},\phi^{(d-\Delta_{UV})}]+\cdots
\end{equation}
where the dots refer to contributions depending on lower order terms $g^{(n<d)}$.

When we do the non-linear PBH transformation, the energy-momentum tensor becomes dependent on the dressed sources \eqref{dressedsources}. The transformation is fully covariant, except for the additional shift of the order $d$ component \eqref{shiftmetric}, that produces an additional contribution to the energy-momentum tensor
\begin{equation}
\delta \vev{T_{\mu\nu}}[\hat g,\hat \phi_{UV}]= -\tau(x) \frac{d}{\kappa^2}\left(\Luv^{d-1}-\Lir^{d-1}\right)e^{-(d-2)\tau} h^{(d)}_{\mu\nu}[\hat g_{\mu\nu},\hat \phi_{UV}].
\end{equation}
As was explained in \cite{deHaro:2000xn}, $h^{(d)}_{\mu\nu}$ equals the variation with respect to the metric of the conformal anomaly, so this implies that the generating functional contains a new term
\begin{equation}
\Gamma\supset \frac{d}{\kappa^2}\left(\Luv^{d-1}-\Lir^{d-1}\right)\int d^d x\,\sqrt{-\hat g}\tau(x)\,\hat \cA_d,
\end{equation}
In agreement with our previous derivation \eqref{anomterm1}. Note however that the variation of the Euler density is zero, so strictly speaking this derivation only captures the contributions to the conformal anomaly that are Weyl-invariant.

The overall coefficient of the Euler density in the new term is proportional to the difference of the $a$-central charges of the UV and IR fixed points \cite{Henningson:1998gx}
\begin{equation}
a_{UV}-a_{IR} \propto \frac{d}{\kappa^2}\left(\Luv^{d-1}-\Lir^{d-1}\right).
\end{equation}
The holographic $a$-theorem of \cite{Freedman:1999gp} implies that $\Luv\geq \Lir$, so indeed the coefficient of the anomalous term is positive. For dimensions larger than four this is a prediction from holography, there is no proof in field theory of this statement. The holographic calculation may be a hint that such a proof actually exists.

For two or four dimensions the result is in agreement with the field theory analysis. Note that the trace of the new contribution to the energy-momentum tensor vanishes $g^{(0)\,\mu\nu}\delta \vev{T_{\mu\nu}}=0$, so the new contribution does not affect to the trace anomaly in the one-point function. The trace anomaly is proportional to the difference of central charges, but it is determined by the trace of the $g^{(d)}_{\mu\nu}$ term
\begin{equation}
g^{\mu\nu}\vev{T_{\mu\nu}}[ g,\phi_{UV}]= \frac{d}{\kappa^2}\left(\Luv^{d-1}-\Lir^{d-1}\right) \cA_d.
\end{equation}
This is the expected result based on physical grounds, by introducing the cutoff $\rhir$ we are discarding the massless degrees of freedom at the IR fixed point, so the anomaly generated by degrees of freedom above the cutoff is proportional to the total anomaly minus the anomaly of the IR CFT.

\section{Spontaneous breaking and Goldstone boson}\label{sec:spont}

When conformal symmetry is spontaneously broken we expect in principle that a massless Goldstone boson associated to dilatations will be present at low energies.\footnote{In $d= 2$ dimensions the analysis becomes more subtle, since IR divergences prevent the appearance of free massless modes, although it is still possible to have massless scalar states in the theory \cite{Coleman:1973ci}.} According to the anomaly matching arguments of \cite{Komargodski:2011vj} its effective action should contain an anomalous term similar to the one we have computed for the spurion field. Although it is encouraging that one should find the same result for the generating functional as in the field theory calculation, there are some caveats in the use of the spurion field that make a separate analysis of the dilaton mode necessary. The caveats are the use of a radial cutoff in order to compute the generating functional in a derivative expansion and the fact that introducing the spurion field alters the boundary values of the metric and other fields; according to the usual dictionary of the AdS/CFT correspondence, the spurion should not be interpreted as a dynamical mode of the dual field theory.

Our goal in this section is to study correlation functions of the energy-momentum tensor and the scalar operator whose expectation value breaks conformal invariance. When the symmetry is broken spontaneously the behavior of these correlators at low momenta is constrained. In particular, a massless pole should appear in the transverse part of mixed tensor-scalar correlator, due to the existence of zero energy states.

In order to compute correlation functions we need to study small fluctuations of the metric and the scalar field. Related works are the calculation of correlation functions \cite{DeWolfe:2000xi,Arutyunov:2000rq,Mueck:2001cy,Papadimitriou:2004rz} in a geometry dual of a particular state in the Coulomb branch of $\cN=4$ SYM \cite{Freedman:1999gk}, that is an example where there is a massless pole that one can identify with the dilaton. In this case the near-horizon geometry is not AdS, but a singular geometry and there is a mass gap in the spectrum. In \cite{Porrati:1999ew} a calculation in a four-dimensional flow between two fixed points was carried out, with the result that at small momenta the two-point function of a probe scalar becomes that of the IR CFT.
A calculation of correlators in a two dimensional flow was also made in \cite{Berg:2002hy}.

Although for a general superpotential we were not able to find the exact solutions to the equations of motion, we can do an expansion in momentum that is possible to solve order by order. In order to capture the low momentum behavior of the correlation functions this expansion is sufficient. The expansion however breaks down very close to the horizon, where regularity or other physical conditions have to be imposed in order to fix the solution.
Fortunately, we can find the near-horizon solution using a different approximation, that in this case is to approximate the near-horizon geometry by $AdS$. There is a region in the geometry where the two approximations overlap, and the matching between the two fixes the form of the correlators at small momenta. In appendix \ref{app:coulomb} we check that this procedure indeed captures the leading low momentum behavior in a geometry dual of $\cN=4$ SYM in the Coulomb branch, where the full solution can be computed analytically.

\subsection{Tensor and scalar fluctuations}

We will solve the equations of motion for the scalar and tensor modes using a low momentum approximation. We will distinguish two regions in the geometry. In the first region, that goes from the boundary to close to the horizon, the solutions are expanded in momentum $q^2$. We solve first for the zero momentum solution and then use it as a starting point to find the next order corrections iteratively. For very low momentum this expansion is valid in the region close to the horizon where the geometry is approximately AdS as long as $q^2 L_{IR}^2 e^{-2r/L_{IR}}\ll 1$. In this way one can construct the two independent solutions corresponding to normalizable and non-normalizable asymptotic behavior at the boundary. In order to compute correlation functions of the dual theory one needs to impose a boundary condition at the horizon that fixes a relation between the coefficient of the two solutions. However, when $r\to -\infty$ the low-momentum expansion will break down, and this defines the second (near-horizon) region.  We take the geometry in the second region to be AdS as a good approximation, and solve for arbitrary values of the momentum $q^2$, imposing the boundary condition that the solutions are ingoing at the horizon.\footnote{This would give retarded correlators. An equivalent condition is to impose a regularity condition at the horizon in Euclidean signature.} For low enough momentum there is an overlap between the first and second regions, so that, to leading order, the low-momentum expansion of the solution found in the second region should coincide with the near-horizon limit of the solution found in the first region. Comparing the two solutions one can then fix the relation between normalizable and non-normalizable modes and use this information to compute the correlation functions of the dual theory.

In the models we are considering the dual of the RG flow is described by Einstein gravity coupled to a scalar field. The equations of motion are
\begin{equation}\label{einsteqs}
R_{MN}=-\partial_M\phi\partial_N\phi-\frac{2}{d-1}g_{MN}V(\phi), \ \ \ \nabla^2 \phi=\frac{\partial V}{\partial \phi}.
\end{equation}
We will adopt Gaussian coordinates for the metric,
\begin{equation}
ds^2=g_{MN}dx^M dx^N=dr^2+g_{\mu\nu}(r,x)dx^\mu dx^\nu.
\end{equation}
The choice $g_{\mu r}=0$, $g_{rr}=1$ corresponds to fixing $d+1$ diffeomorphisms up to transformations of the form
\begin{equation}
\delta g_{MN}=\nabla_{(M} \xi_{N)}=0, \quad {\rm if} \ M=r \; {\rm and / or} \; N=r.
\end{equation}
For the background metric $g_{\mu\nu}=e^{2A}\eta_{\mu\nu}$, we can distinguish between $d$-dimensional diffeomorphisms
\begin{equation}\label{diff1}
\xi_r=0, \ \ \xi_\mu=e^{2 A} \sigma_\mu(x),
\end{equation}
and translations in the radial direction
\begin{equation}\label{gaugetr}
\xi_r=-\sigma(x), \ \ \xi_\mu=\partial_\mu\sigma(x) e^{2 A} \int^r dr' e^{-2 A(r')}.
\end{equation}
For $\sigma$ a constant, these transformations become scale transformations in the $AdS$ boundary, so it is natural to associate them with dilatations in the dual field theory.

We will expand the equations to linear order in fluctuations
\begin{equation}
g_{\mu\nu}=e^{2A}(\eta_{\mu\nu}+h_{\mu\nu}), \ \ \phi=\phi_0+\varphi,
\end{equation}
and use the flat metric to raise and contract indices. For instance, the trace of the metric fluctuation is defined as $h=h^\mu_\mu=\eta^{\mu\nu}h_{\mu\nu}$.
We will use an expansion in Fourier modes. Each mode of the metric admits the following decomposition in transverse and longitudinal parts, traceless and not traceless
\begin{equation}\label{hdec}
h_{\mu\nu}=h_{\mu\nu}^{TT}+h_{\mu\nu}^{TL}+h_{\mu\nu}^T+h_{\mu\nu}^L.
\end{equation}
More explicitly, we define the projectors
\begin{align}\label{projector}
\notag &P_T^{\mu\nu}=q^2\eta^{\mu\nu}-q^\mu q^\nu,\ \ P_L^{\mu\nu}=q^\mu q^\nu\\
&\Pi^{\mu\nu,\alpha\beta}=P_T^{\mu\alpha}P_T^{\nu\beta}-\frac{1}{d-1}P_T^{\mu\nu}P_T^{\alpha\beta},
\end{align}
so that
\begin{align}\label{qtens}
\notag & h_{\mu\nu}^{TT}=\frac{1}{(q^2)^2}\Pi_{\mu\nu}^{\ \ \ \alpha\beta}h_{\alpha\beta},\\
\notag & h_{\mu\nu}^{TL}=\frac{1}{(q^2)^2}\left(P_{T \mu}^{\ \ \alpha}P_{L \nu}^{\ \ \beta}+P_{T \nu}^{\ \ \beta}P_{L \mu}^{\ \ \alpha} \right)h_{\alpha\beta},\\
\notag & h_{\mu\nu}^T=\frac{1}{(d-1)(q^2)^2}P_{T\ \mu\nu}P_T^{\alpha\beta} h_{\alpha\beta},\\
 & h_{\mu\nu}^L=\frac{1}{(q^2)^2}P_{L\ \mu}^{\ \ \alpha}P_{L \ \nu}^{\ \ \beta} h_{\alpha\beta}.
\end{align}
For convenience, we will also define $h=h_T+h_L$, where
\begin{equation}
h_T=\eta^{\mu\nu}h_{\mu\nu}^T=\frac{1}{q^2}P_T^{\alpha\beta}h_{\alpha\beta}, \ \ h_L=\eta^{\mu\nu}h_{\mu\nu}^L=\frac{1}{q^2}P_L^{\alpha\beta}h_{\alpha\beta}.
\end{equation}
The details of the derivation of the equations of motion can be found in Appendix \ref{app:eoms}. Here we quote the results for the two modes we will analyze in the following:
\begin{itemize}
\item Tensor fluctuation:
\begin{equation}\label{eqTT}
0={h^{TT}_{\mu\nu}}''+dA'{h^{TT}_{\mu\nu}}'-q^2e^{-2A}{h^{TT}_{\mu\nu}}.
\end{equation}
\item Scalar fluctuations:\\
There are two constraints for the trace components of the metric
\begin{align}
\label{eqA} &0= (d-1)A'h'+2\partial V\varphi -2\phi_0'\varphi'-q^2 e^{-2A} h_T,\\
\label{consT} &0=h_T'+2\phi_0'\varphi,
\end{align}
and, defining $h'=e^{-2A}H$, two second order equations for the metric and the scalar fluctuations:
\begin{align}
\label{scalareqs1} &0= e^{-2A}H'+\frac{4}{d-1}\partial V \varphi+4\phi_0'\varphi',\\
\label{scalareqs2} &0=\varphi''+dA'\varphi'-\partial^2 V\varphi-q^2 e^{-2A}\varphi+\frac{\phi_0'}{2}e^{-2A}H.
\end{align}
\end{itemize}
The equation of motion for the vector fluctuation is simply ${h^{TL}}'=0$. This fluctuation is not dynamical and can be set to zero using the residual diffeomorphism invariance \eqref{diff1}.

\subsection{Tensor fluctuation}

We start with the simpler case of a tensor fluctuation, given by \eqref{eqTT}, with $h_{\mu\nu}^{TT}\equiv \frac{1}{(q^2)^2}\Pi_{\mu\nu}^{\ \ \alpha\beta}\varepsilon_{\alpha\beta} h^{TT}$, where $\varepsilon_{\mu\nu}$ is an arbitrary symmetric tensor. We will also define $Q\equiv \sqrt{-q^2}=\sqrt{\omega^2-\vec{q}^2}$. The analysis can be extended to Euclidean signature by considering spacelike momenta $Q=i\sqrt{-\vec{q}^2}$.

In general, we cannot solve \eqref{eqTT} exactly but we can solve it perturbatively for small $Q^2$
\begin{equation}\label{tenssol}
h^{TT}=h_{b}^{TT}+h_{N}^{TT}\int dr e^{-dA}+O(Q^2),
\end{equation}
where $h_{b}^{TT}$ and $h_{N}^{TT}$ are the coefficients of the non-normalizable and normalizable modes, respectively. The value of $h_{b}^{TT}$ determines the tensor part of the boundary metric. They do not depend on the radial coordinate but may depend on $q$. One can solve recursively for the next orders
\begin{equation}
h^{TT}=\sum_{n\geq 0} Q^{2n} h^{TT}_{(n)},
\end{equation}
where $H^{(0)}=H_b^T$ and the recursive equation is
\begin{equation}
{h^{TT}_{(n)}}''+dA'{h^{TT}_{(n)}}'=-e^{-2A}Q^2 h^{TT}_{(n-1)}
\end{equation}
The solution to this equation takes the form
\begin{equation}
 h^{TT}_{(n)}=-\int dr e^{-dA} \int dr e^{(d-2)A}Q^2 h^{TT}_{(n-1)}.
\end{equation}
One should actually keep terms up to order $O(Q^d)$ (for $d$ even), since the normalizable term is of this order.

If $Q^2$ is small enough this solution can be a good approximation even close to the horizon where the geometry is approximately AdS. Assuming $(Q L_{IR})^2 e^{-2r/\Lir}\ll 1$,
\begin{eqnarray}\label{hb}
\nonumber  h^{TT} &\simeq& h_b^{TT}\left[1+O(Q^2 e^{-2r/\Lir})\right] \\
                    &+& e^{-dr/\Lir}\left[-\frac{\Lir}{d}h_{N}^{TT} +\left(\frac{\Lir Q}{2(d-2)} \right)^dh_b^{TT}+O(Q^2e^{-2r/\Lir})\right].
\end{eqnarray}
This expression is valid for $d>2$ and even. The perturbative solution breaks down very close to the horizon at $r\to -\infty$ since the factor $e^{-2A}$ that multiplies $Q^2$ in the equation becomes very large. Therefore we cannot use this solution in order to impose boundary conditions in the IR. For that we will have to solve the equation of motion in the near horizon region, where it takes the form
\begin{equation}
0={h^{TT}_h}''+\frac{d}{\Lir}{h^{TT}_h}'-e^{-2r/\Lir}q^2 h^{TT}_h.
\end{equation}
The solutions to this equation are Bessel functions. There are several possible choices of boundary conditions at the horizon. We impose ingoing boundary conditions at the horizon, following the prescription of \cite{Son:2002sd}. This condition basically means that the solution behaves as an ingoing plane-wave in the vicinity of the horizon. Note that if the solution is continued analytically to Euclidean signature, this choice corresponds to fixing the exponential behavior of the solution, in such a way that the solution is regular.

The solution satisfying ingoing boundary conditions at the horizon takes the form
\begin{equation}\label{hh}
h^{TT}_h=C\left(\Lir Q e^{-r/\Lir}\right)^{d/2} H^{(1)}_{\frac{d}{2}}\left(\Lir Q e^{-r/\Lir}\right).
\end{equation}
where $H_\frac{d}{2}^{(1)}(x)$ is the Hankel function of the first kind.

In the low momentum limit there is a region in the geometry where both the expansion in momentum of the solution $h^{TT}$ and the AdS approximation $h^{TT}_h$ are valid. This will be when $r\to-\infty$ but $\Lir Q e^{-r/\Lir}\ll 1$. In this case we should be able to match the leading contributions of both solutions \eqref{hb} and \eqref{hh}. In the case at hand, for $d$ even dimensions, the solution \eqref{hb} in the limit $r\rightarrow - \infty$ takes the form
\begin{equation}
h^{TT}\simeq h_{b}^{TT}-\frac{\Lir}{d}h_{N}^{TT} e^{-dr/\Lir}+\cdots
\end{equation}
while the solution \eqref{hh} in the limit $\Lir Q e^{-r/\Lir}\ll 1$ is
\begin{equation}
h^{TT}_h \simeq C\left(-\frac{i a_d}{\pi}+\cdots +\frac{ib_d}{\pi}(\Lir Q)^d\log(\Lir Q)^2e^{-dr/\Lir}+\cdots \right)
\end{equation}
where for $d=2,4,6$, $a_d=2,4,6$ and $b_d=1/2, 1/8,1/48$. There are analytic terms proportional to $(\Lir Q)^d$, as the term proportional to $h_{b}^{TT}$ in \eqref{hb}, but for $d\geq 4$ those will only introduce contact terms in the correlation function and we can ignore them.

In both cases we have identified the leading terms of the normalizable and non-normalizable solutions. Note that the non-normalizable solution can have contributions that are less suppressed than the normalizable solution when $r\to\-\infty$. The matching of the two solutions determines
\begin{equation}
C=\frac{i\pi}{a_d} h_{b}^{TT}, \ \ h_{N}^{TT}=\frac{d}{\Lir}\frac{b_d}{a_d}(\Lir Q)^d\log(\Lir Q)^2h_{b}^{TT}.
\end{equation}
Then, the ration between the non-normalizable and normalizable solutions is
\begin{equation}\label{irGtensor}
\frac{h_{N}^{TT}}{ h_{b}^{TT}}=\frac{d}{\Lir}\frac{b_d}{a_d}(\Lir Q)^d\log(\Lir Q)^2,
\end{equation}
The dependence on the momentum is in agreement with the two-point function of the energy-momentum tensor in a CFT.

In terms of the boundary value of the metric $h_{b\,\mu\nu}$, we can write the fluctuation as
\begin{equation}
h^{TT}_{\mu\nu}=\frac{1}{(q^2)}\Pi_{\mu\nu}^{\ \ \ \alpha\beta}h_{b\,\alpha\beta}\left[1+G^{TT}(q)e^{-dr/\Luv}+\cdots\right],
\end{equation}
where, at low momentum
\begin{equation}\label{gtt}
G^{TT}(q)\simeq -\frac{\Luv}{\Lir}\frac{b_d}{a_d}(\Lir Q)^d\log(\Lir Q)^2.
\end{equation}

\subsection{Scalar fluctuation}

A scalar fluctuation involves both the scalar field and the scalar components of the metric, that are coupled through (\ref{scalareqs1},\ref{scalareqs2}). We will solve the equations using the same approximations as for the tensor mode, but this is a more complicated case because we are dealing with coupled equations. In order to solve the system we will first write a single equation for the scalar fluctuation $\varphi$, solve it, and then check the remaining equations, including the constraints. We will find that there are only two independent solutions. The relation between the coefficients of the two independent solutions, which will determine the correlator of the dual scalar operator, will be fixed by a boundary condition at the horizon.

We can decouple the scalar field by multiplying \eqref{scalareqs2} by $e^{2A}/\phi_0'$, taking the derivative once with respect to the radial coordinate, substituting $H'$ by the expression obtained from \eqref{scalareqs1} and multiplying by $\phi_0' e^{-2A}$. The result is a third order equation for $\varphi$, for which there are three solutions. In order to get a simplified expression one can use the first order equations of the background scalar and metric:
\begin{align}
\notag &A'=\frac{W}{d-1}, \ \ \phi_0'=-\partial W,  \ \ A''=-\frac{(\partial W)^2}{d-1},\\ & \phi_0''=\partial^2 W\partial W, \ \ \phi_0'''=-\partial^2 W(\partial W)^2-(\partial^2 W)^2\partial W .
\end{align}
We will also change variables from the radial coordinate $r$ to the background scalar field $\phi_0$, using that $\partial_r=\phi_0'\partial=-\partial W\partial$, where $\partial$ is now a derivative with respect to the background field $\phi_0$. In addition, we substitute the potential and its derivative by expressions in terms of the superpotential \eqref{potential}. The resulting equation has the form
\begin{equation}
\partial^3 \varphi+\alpha_2\partial^2\varphi+\alpha_1\partial\varphi+\alpha_0\varphi=0.
\end{equation}
Where
\begin{align}
\alpha_2 &=-\frac{(d+2) W}{(d-1) \partial W}+\frac{2 \partial^2 W}{\partial W},\\
\alpha_1 &= -\frac{q^2 e^{-2 A}}{(\partial W)^2}+\frac{d^2 W \partial^2 W-3 d W \partial^2 W+2 d W^2+2 W \partial^2 W}{(d-1)^2
   (\partial W)^2}-\frac{(\partial^2 W)^2}{(\partial W)^2}-2,\\
\notag \alpha_0 &=\frac{q^2 e^{-2 A} \partial^2 W}{(\partial W)^3}+\frac{-\frac{\partial^3 W \left(d^2 (-W)-d W+2 W\right)}{(\partial W)^2}-\frac{2 d W^2   \partial^2 W}{(\partial W)^3}}{(d-1)^2}\\
&-\frac{d W (\partial^2 W)^2-2 W (\partial^2 W)^2}{(d-1)
   (\partial W)^3}-\frac{\partial^4 W}{\partial W}+\frac{(\partial^2 W)^3}{(\partial W)^3}+\frac{2 \partial^2 W}{\partial W}-\frac{2 \partial^3 W \partial^2 W}{(\partial W)^2}.
\end{align}
The form of the equation is quite involved, but we can simplify things by factorizing out one solution. This is possible thanks to the residual diffeomorphism \eqref{gaugetr}, under which the scalar field transforms as
\begin{equation}
\delta \phi= \phi_0' \xi_r =-\sigma(x)\partial W.
\end{equation}
The metric is also affected by this transformation
\begin{equation}\label{metg}
\delta g_{\mu\nu}= \partial_\mu \xi_\nu+ \partial_\nu \xi_\mu+\xi_r\partial_r g_{\mu\nu}=2 e^{2 A}\left( \partial_\mu\partial_\nu\sigma(x)\int^r dr' e^{-2 A}-\frac{W}{d-1}\eta_{\mu\nu}\sigma(x)\right).
\end{equation}
The \emph{gauge mode} $\varphi_g=\partial W$ is a solution to the equation of motion. This fact enables us to factorize it as follows
\begin{equation}\label{eqscalar}
(\partial^2+a_1\partial +a_0)\left(\partial-\frac{\partial^2 W}{\partial W }\right)\varphi=0,
\end{equation}
where
\begin{align}
\notag &a_1=3\frac{\partial^2 W}{\partial W}-\frac{(d+2) W}{ (d-1) \partial W},\\
&a_0=-\frac{q^2 e^{-2 A}}{(\partial W)^2}-\frac{4}{d-1}\frac{ W\partial^2 W}{ (\partial W)^2}+\frac{2 d }{ (d-1)^2}\frac{W^2}{(\partial W)^2}+\frac{2 \partial^3 W}{\partial W}-2.
\end{align}
The full solution is a superposition of the gauge solutions plus two more solutions
\begin{equation}
\varphi=C_g \partial W+C_1\varphi_1+C_2\varphi_2.
\end{equation}
Where $C_g$, $C_1$ and $C_2$ and arbitrary coefficients. Note that the gauge mode is a valid solution for any value of the momentum and of the radial coordinate and that it vanishes at the boundary in the way expected for a normalizable solution $\varphi \sim \phi_0$. The transformation of the metric \eqref{metg} on the other hand does not vanish at the boundary. Then, imposing a Dirichlet boundary condition on the metric will fix this mode, we will do this explicitly below but first we will fix the boundary conditions at the horizon.

We proceed to find the two other solutions $\varphi_1$ and $\varphi_2$. In order to simplify things further, we define
\begin{equation}\label{gaugeinvG}
\left(\partial-\frac{\partial^2 W}{\partial W }\right)\varphi=\frac{W}{(\partial W)^2}e^{-dA} G.
\end{equation}
Then, $G$ satisfies the following second order equation
\begin{equation}\label{eqG}
\partial^2 G+\partial B \partial G-q^2\frac{e^{-2A}}{(\partial W)^2}G=0,
\end{equation}
where
\begin{equation}
B=2 \log(W)-\log(\partial W)-(d-2)A.
\end{equation}
In general, we cannot solve \eqref{eqG} exactly but we can use the matching procedure in the same way as for the tensor components of the metric. The function $G$ is explicitly gauge-invariant, it coincides up to a momentum-independent factor with the variable $R$ defined in section 4 of \cite{Bianchi:2001de} when the $g_{rr}$ component of the metric is fixed.

First we solve \eqref{eqG} perturbatively for small $q^2$. The solution takes the form
\begin{eqnarray}\label{q2sol}
\nonumber  G &=& C_2\left[ 1+q^2\int d\phi _0 e^{-B} \int d\phi _0 \frac{e^{B-2A}}{(\partial W)^2} +\mathcal{O}(q^4) \right] \\
   &+& C_1 \left[ \int d\phi _0 e^{-B} + q^2 \int d\phi _0 e^{-B} \int d\phi _0 \frac{e^{B-2A}}{(\partial W)^2} \int d\phi _0 e^{-B} +\mathcal{O}(q^4) \right]
\end{eqnarray}
Then we solve \eqref{eqG} in the near horizon region, where it takes the form
\begin{equation}
\partial ^2 G + \frac{\frac{d-2}{\lambda}-1}{\phi_0 - \phi _m} \partial G +\left(\frac{L_{IR}Q}{\lambda}\right)^2 (\phi_0 - \phi _m)^{\frac{2}{\lambda} -2} G = 0
\end{equation}
The solutions to this equation are Bessel functions. We will pick up a solution satisfying ingoing boundary conditions at the horizon (regularity in the Euclidean)
\begin{eqnarray}\label{horizonSol}
\nonumber G &=& C_h \left( L_{IR}Q (\phi_0 - \phi _m)^{\frac{1}{\lambda}} \right) ^{-\nu} H^{(1)}_{\nu} \left( L_{IR}Q (\phi_0 - \phi _m)^{\frac{1}{\lambda}} \right) \\
\nu &\equiv& \frac{d-2-2\lambda}{2}
\end{eqnarray}
where $C_h$ is an arbitrary coefficient and we have defined
\begin{equation}
\lambda = d- \Delta _{IR}.
\end{equation}
Note that the relation between $\varphi$ and $G$ \eqref{gaugeinvG} is such that close to the horizon $\varphi$ has different power-like dependence on $(\phi_0-\phi_m)$, but the exponential behavior is the same. Then, the ingoing condition on $G$ is equivalent to an ingoing condition for $\varphi$. Note that the gauge mode does not contribute to the exponential behavior since $\partial W \sim (\phi_0-\phi_m)$, so it is not affected by the ingoing boundary condition. One can systematically find corrections to this solution by expanding the coefficients of the differential equation in powers of $(\phi-\phi_m)$ and solving order by order.

Expanding the perturbative solution \eqref{q2sol} around the horizon $\phi_0 = \phi _m$ we have
\begin{eqnarray}\label{Gboundary}
G & \simeq & C_2 + C_1 \frac{L_{IR}\lambda ^2}{(d-1)^2(2\lambda-d+2)}\left(\phi_0 - \phi _m\right)^{\frac{2\lambda -d+2}{\lambda}}
\end{eqnarray}
Expanding the near horizon solution \eqref{horizonSol} in small momenta
\begin{equation}
G\simeq C_h \left[ a + b \left( L_{IR}Q (\phi_0 - \phi _m)^{\frac{1}{\lambda}} \right) ^{2\lambda -d+2} \right]
\end{equation}
where
\begin{equation}
a=\frac{2^{-\nu}(1+i \cot(\pi \nu))}{\Gamma (\nu+1)}  , \qquad b=-\frac{i 2^{\nu} \Gamma (\nu)}{\pi}
\end{equation}
As we show in Appendix \ref{app:matching}, higher order corrections to the near-horizon solution do not affect the leading terms in the low momentum expansion. Matching the two solutions fixes
\begin{equation}
\frac{C_1}{C_2} = \frac{b}{a} \left(L_{IR}Q\right)^{2\lambda-d+2} = \frac{b}{a} \left(L_{IR}Q\right)^{d-2\Delta_{IR}+2}
\end{equation}

Near the boundary the solution for $G$ takes the same form as \eqref{Gboundary} with the replacements $L_{IR}\rightarrow L_{UV}$, $\lambda = \Delta _{UV}$ and $\phi_0 - \phi _m \rightarrow \phi_0$ (see \eqref{superpot}-\eqref{superpot2}). Solving \eqref{gaugeinvG} for $\varphi$ and including the gauge solution $\varphi=C_g \partial W$, we can then write its asymptotic form
\begin{align}
\notag \varphi &= \phi _0 \left[ C_g \frac{\Delta_{UV}}{L_{UV}} + \frac{C_1}{ (\Luv Q)^2} \frac{\Delta_{UV}}{2(d-1)(2\Delta_{UV} -d +2)} \left(L_{UV}Q \phi _0 ^ {\frac{1}{\Delta_{UV}}}\right)^2 \right]\\
&+\phi_0 ^{\frac{d}{\Delta_{UV}}-1} \left[C_2 L_{UV} \frac{d-1}{\Delta_{UV}(d-2\Delta_{UV})} \right]. \label{scalarSol}
\end{align}
In order to simplify the expressions we will define the leading order coefficients of the expansion close to the boundary $\phi_0\to 0$ as:
\begin{equation}
\varphi \simeq \varphi_N \phi_0+C_N \phi_0^{1+\frac{2}{\Delta_{UV}}}+\varphi_b\phi_0^{\frac{d}{\Delta_{UV}}-1}.
\end{equation}
Where
\begin{align}
\varphi_N= \frac{\Delta_{UV}}{L_{UV}}C_g , \ \ C_N =\frac{\Delta_{UV}C_1}{2(d-1)(2\Delta_{UV} -d +2)}, \ \ \varphi_b=  \frac{(d-1)L_{UV} C_2}{\Delta_{UV}(d-2\Delta_{UV})}.
\end{align}
The metric fluctuations $h$ and $h_T$ can be computed from \eqref{scalareqs1} and \eqref{consT} respectively. The expansion of the transverse component close to the  boundary is, to leading order
\begin{align}\label{htexp}
\notag h_T &= h_{T\, b}-2C_g(W-W(0))-2\int_0^{\phi_0} d\phi_0\,(\varphi-C_g \partial W)\\ &\simeq  h_{T\, b}-\varphi_N\phi_0^2-\frac{C_N}{\Delta_{UV}+1}\phi_0^{2+\frac{2}{\Delta_{UV}}}-\frac{2\Delta_{UV}}{d}\varphi_b\phi_0^{\frac{d}{\Delta_{UV}}}.
\end{align}
Here $h_{T\, b}$ is the boundary value of the metric, that we fix by imposing Dirichlet boundary conditions. This determines the lower limit of integration, that should be such that no additional contributions change the value of $h_{T\, b}$.

The expansion of the full trace is
\begin{equation}\label{hexp}
h \simeq h_{ b}-\frac{d}{d-1}\varphi_N\phi_0^2-C_N\frac{d(\Delta_{UV}+1)-2}{(d-1)(\Delta_{UV}+1)}\phi_0^{2+\frac{2}{\Delta_{UV}}}+\frac{4\Delta_{UV}(\Delta_{UV}-d)}{d(d-1)}\varphi_b\phi_0^{\frac{d}{\Delta_{UV}}}
\end{equation}
Plugging \eqref{htexp} and \eqref{hexp} into \eqref{eqA}, we find the following condition
\begin{equation}\label{cond}
\varphi_N =(d-2\Delta_{UV}-2)\frac{C_N}{(Q\Luv)^2}.
\end{equation}
This relation implies that the close to the boundary the solution for the scalar field has the same radial dependence as the solutions in the decoupled system, where the background scalar is set to zero and the geometry is pure $AdS$.

In terms of $C_g$:
\begin{equation}\label{Cg}
C_g = -\frac{L_{UV}}{(d-1)} \frac{C_1}{(\Luv Q)^2} = -\frac{L_{IR}^2}{(d-1)\Luv} \frac{b}{a} \left( L_{IR}Q \right)^{d-2\Delta_{IR}} C_2
\end{equation}
Summarizing, if we define the coefficients of the boundary expansion of the scalar field and the scalar component of the metric as
\begin{eqnarray}
  \varphi &=& \varphi_b \phi_0^{\frac{d}{\Delta_{UV}}-1}+\varphi_N\phi_0\cdots \\
  h &=& h_b + h_N \phi_0^{\frac{d}{\Delta_{UV}}}\cdots
\end{eqnarray}
we find that
\begin{equation}
\varphi_N=C_g \frac{\Delta_{UV}}{L_{UV}}= G_s(q)\varphi_b,
\end{equation}
where
\begin{equation}\label{gs}
G_s(q)=\frac{b}{a}\frac{1}{2\Delta_{UV}-d}\frac{\Lir^2}{\Luv^2}  \left( L_{IR}Q \right)^{d-2\Delta_{IR}}.
\end{equation}
For the metric we find, using \eqref{consT} and \eqref{eqtracea}
\begin{eqnarray}
  h_{N T} &=& G^T (q) \varphi_b, \\
  h_{N L} &=& G^L (q) \varphi_b,
\end{eqnarray}
where
\begin{eqnarray}
  G^T (q) &=& -2 \frac{\Delta_{UV}}{d}, \\
  G^L (q) &=& \frac{2\Delta_{UV}(2\Delta_{UV}-d-1)}{d(d-1)} .
\end{eqnarray}

We have checked that the equations for the gauge-invariant variable $R$ in \cite{Bianchi:2001de} and the equation for $G$ are equivalent and lead to the same results in the Coulomb branch geometry. The conventions  are such that to map to our results one should change the normalization of the superpotential and the scalar field as $W\to -W/2$, $\phi\to\phi/\sqrt{2}$. In this case, the equation that determines the normalizable part of $\phi$, (4.10) in \cite{Bianchi:2001de}, is proportional to $\partial G$ and then, using the solutions for $G$ that we derive, one can see that the overall factor $1/q^2$ is canceled, leaving a $q^{2\Delta_{IR}-d}$ dependence. In the Coulomb branch $\partial G$ goes as $O(q^0)$ to leading order and the $1/q^2$ factor eventually introduces a pole in the correlation function.

\subsection{One and two-point functions}
In the previous sections we have found the solutions of the linearized equations of motion for the metric and the scalar fields in a low-momentum approximation. We will use them to compute correlation functions applying the usual holographic dictionary. The classical on-shell action for the fluctuations equals the generating functional of the dual field theory. The coefficients of the non-normalizable modes $h_{b\,\mu\nu}$ and $\varphi_b$ are proportional to sources in the dual field theory for the energy-momentum tensor and the scalar operator $\cO$. One can then compute correlation functions taking the variation of the on-shell action with respect to these coefficients. In order to have a well-defined variation one has first to ensure that the action is finite, this requires adding counterterms depending on the fields evaluated at a radial cutoff that acts as a regulator.

We will now compute the on-shell action for the fluctuations up to quadratic order (this is all is needed for one- and two-point functions), adding the necessary counterterms. Because of the low-momentum expansion we will need to worry only about counterterms with no derivatives of the fields, although a full treatment will require to take into account also counterterms with derivatives.

In order to compute the correlation functions of the scalar operator and the energy-momentum tensor, we need to evaluate the on-shell action for the linearized fluctuations. We can use the results of \cite{Mueck:2001cy,Bianchi:2001de,Papadimitriou:2004rz} adapted to our case. The on-shell action to quadratic order in the fluctuations is
\begin{equation}
S[g_{\mu\nu}+e^{2A}h_{\mu\nu},\phi_0+\varphi]=S[g_{\mu\nu},\phi_0]+\delta S\left[g_{\mu\nu}+\frac{1}{2}e^{2A}h_{\mu\nu},\phi_0+\frac{1}{2}\varphi ;e^{2A}h_{\mu\nu},\varphi \right],
\end{equation}
where the variation of the on-shell action coincides with \eqref{varonshell} but now the only contribution is evaluated at the boundary and we are using the Gaussian coordinate $r$, so the extrinsic curvature is $K_{\mu\nu}=\frac{1}{2}\partial_r g_{\mu\nu}$ and the variation is
\begin{equation}
\delta S[g_{\mu\nu},\phi_0;e^{2A}h_{\mu\nu},\varphi ]=\lim_{r\to \infty} -\frac{1}{\kappa^2}\int d^d x\sqrt{-g}\left[\frac{1}{2}(K^{\mu\nu}-K g^{\mu\nu})e^{2A}h_{\mu\nu}+\phi_0' \varphi \right].
\end{equation}
We will implicitly assume the $r\to \infty$ limit. In order for the action to be well defined we will have to add counterterms that cancel the divergent pieces. The counterterms take the form \cite{Papadimitriou:2004rz}\footnote{This choice of counterterms corresponds to a particular renormalization scheme, as explained in \cite{Papadimitriou:2004rz}. In this case the energy density of the vacuum will be zero for the particular flows we are considering. In general any function which is a solution to the differential equation that relates the potential to the superpotential in \eqref{potential} can be used as counterterm, for superpotentials that describe explicit breaking. We want to thank Ioannis Papadimitriou for explaining this point to us and Kostas Skenderis for further clarifications.}
\begin{equation}
S_{ct}=-\frac{1}{\kappa^2}\int d^d x\sqrt{-g}\left[\frac{d-1}{\Luv}+\frac{d-\Delta_{UV}}{2\Luv}\phi^2\right]+\cdots.
\end{equation}
The first term cancels the volume divergence of the on-shell action, while the second term is necessary in order to cancel divergences from the scalar action. The reason is also explained in \cite{Papadimitriou:2004rz}, the counterterm should be of the form of a superpotential $W$ for a flow that breaks explicitly the symmetry, expanding this superpotential to quadratic order in the scalar field one finds a counterterm of the form above. Note that the superpotential that determines the flow, $W$, has a different quadratic term because it describes spontaneous breaking.

For our analysis it is enough to consider counterterms involving no derivatives. In order to compute corrections that are of higher order in the low momentum expansion, we will need to add more counterterms depending on the curvature of the background metric and derivatives of the scalar field.

To linear order in the fluctuations, the finite contributions in the $r\to \infty$ or, equivalently, the $\phi_0\to 0$ limit are
\begin{equation}
S_{(1)}=-\frac{d-2\Delta_{UV}}{\kappa^2\Luv}\int d^d x\, e^{dA}\phi_0 \varphi  =-\frac{d-2\Delta_{UV}}{\kappa^2\Luv}\int d^d x\, \varphi_b.
\end{equation}
This shows that the one-point function of the scalar field is a nonzero constant. Here $\varphi_b$ is dimensionless, while the product of the source for the scalar field $J_b$ times the expectation value $\vev{\cO}$ has dimension $d$. Then, $\varphi_b$ is related to the actual source by a factor $\varphi_b=M^{\Delta_{UV}-d}J_b$, where $M$ is some mass scale. The expectation value of the scalar operator is
\begin{equation}
\vev{\cO}=\frac{\delta S^{(1)}}{\delta J_b}=-\frac{d-2\Delta_{UV}}{\kappa^2\Luv}M^{\Delta_{UV}-d}.
\end{equation}
We can use this expression to fix the relation between the source $J_b$ and the boundary value of the scalar field $\varphi_b$ as
\begin{equation}\label{sourceJ}
\varphi_b=\frac{\kappa^2\Luv}{2\Delta_{UV}-d}\vev{\cO}J_b.
\end{equation}

The on-shell action to quadratic order is, in our gauge
\begin{align}
\notag S_{(2)} &=-\frac{1}{\kappa^2}\int d^d x\sqrt{-g}\left[\frac{1}{8}{ h_{\mu\nu} h^{\mu\nu}}'-\frac{1}{8}h^\nu_\nu{ h^\mu_\mu}' +\frac{1}{2}\varphi\varphi'+\frac{d-\Delta_{UV}}{2\Luv}\varphi^2\right.\\
&\left.+\left(\frac{1}{4}\phi_0'+\frac{d-\Delta_{UV}}{2\Luv}\phi_0\right) h^\mu_\mu \varphi\right].\label{regaction}
\end{align}
where indices are raised with the flat metric $\eta_{\mu\nu}$. Using the background scalar field $\phi_0$ as radial coordinate, the on-shell action becomes
\begin{align}
\notag S_{(2)} &=-\frac{1}{\kappa^2}\int d^d x\, e^{dA}\partial W\left[-\frac{1}{8}h_{\mu\nu}{ \partial h^{\mu\nu}}+\frac{1}{8}h^\nu_\nu{ \partial h^\mu_\mu} -\frac{1}{2} \varphi \partial \varphi + \frac{d-\Delta_{UV}}{2\Luv\partial W}\varphi^2\right.\\
&\left.+\left(\frac{1}{4}+\frac{d-2\Delta_{UV}}{2\Luv}\frac{\phi_0}{\partial W}\right) h^\mu_\mu \varphi\right].
  \label{regaction2}
\end{align}
Expanding close to the boundary as
\begin{equation}
\varphi=\varphi_b \phi_0^{\frac{d}{\Delta_{UV}}-1}+\varphi_N \phi_0+\cdots, \ \ h_{\mu\nu}=h_{b\ \mu\nu}+h_{N\ \mu\nu}\phi_0^{\frac{d}{\Delta_{UV}}}\cdots,
\end{equation}
One can check that the divergent contributions to the action vanish:
\begin{align}
S_{(2)} &=-\frac{1}{\Luv\kappa^2}\int d^d x\, \phi_0^{1-\frac{d}{\Delta_{UV}}}\left[-\frac{d-\Delta_{UV}}{2} \phi_0^{2\frac{d}{\Delta_{UV}}-1}\varphi_b^2+\frac{d-\Delta_{UV}}{2} \phi_0^{2\frac{d}{\Delta_{UV}}-1}\varphi_b^2\right]=0.
\label{regaction3}
\end{align}
The first contribution comes from the term $\propto \varphi \partial\varphi $, while the second contribution comes from the term $\propto \varphi^2$ in \eqref{regaction2}.

The finite contributions to the action are
\begin{align}
S_{(2)} &=-\frac{1}{\kappa^2\Luv}\int d^d x\, \left[-\frac{d}{8}h_N^{\mu\nu}h_{b\,\mu\nu}+\frac{d}{8} h^{\ \mu}_{N\  \mu} h^{\ \nu}_{b\ \nu}+\frac{d-2\Delta_{UV}}{2}\varphi_N\varphi_b+\frac{2d-3\Delta_{UV}}{4} h^{\ \mu}_{b\ \mu} \varphi_b\right].
  \label{regaction4}
\end{align}

Using the decomposition \eqref{hdec} and \eqref{qtens} and using that $h^{TL}_{\mu\nu}=0$,\footnote{As we commented before, this is always possible using residual diffeomorphisms.} we can write the action as
\begin{align}
\notag S_{(2)} &=-\frac{1}{\kappa^2\Luv}\int \frac{d^d q}{(2\pi)^d} \left[-\frac{d}{8}\left(h^{TT}_{N,\mu\nu}\frac{1}{(q^2)^2}\Pi^{\mu\nu,\alpha\beta}h_{b\ \alpha\beta}\right.\right.
\\ \notag &\left. +h^T_{N\,\mu\nu}\frac{1}{(q^2)^2(d-1)} P_T^{\mu\nu}P_T^{\alpha\beta}h_{b\ \alpha\beta} +h^L_{N\,\mu\nu}\frac{1}{(q^2)^2} P_L^{\mu\alpha}P_L^{\nu\beta}h_{b\ \alpha\beta}\right)\\
&\left. +\frac{d}{8} h^{\ \mu}_{N \ \ \mu} h^{\ \nu}_{b \ \ \nu}+\frac{d-2\Delta_{UV}}{2}\varphi_N\varphi_b+\frac{2d-3\Delta_{UV}}{4} h^{\ \mu}_{b \ \ \mu} \varphi_b\right].\label{regaction5}
\end{align}
The solution we have found depends on the boundary sources as
\begin{align}
& h^{TT}_{N,\mu\nu}=\frac{1}{(q^2)^2}\Pi^{\mu\nu,\alpha\beta}G^{TT}(q)h_{b\ \alpha\beta},\\
& h_{N\,T}=\frac{1}{q^2} P_T^{\mu\nu}h_{N\,\mu\nu}=G^T(q) \varphi_b,\\
& h_{N\,L}=\frac{1}{q^2} P_L^{\mu\nu}h_{N\,\mu\nu}=G^L(q) \varphi_b,\\
& \varphi_N=G_s(q)\varphi_b.
\end{align}
Therefore, using that $\frac{1}{q^2}P_T^{\alpha\beta}+\frac{1}{q^2}P_L^{\alpha\beta}=\eta^{\alpha\beta}$,
\begin{align}
\notag S_{(2)} &=-\frac{1}{\kappa^2\Luv}\int \frac{d^d q}{(2\pi)^d} \left[-\frac{d}{8}h_{b,\mu\nu}\frac{1}{(q^2)^2}\Pi^{\mu\nu,\alpha\beta}G^{TT}(q)h_{b\ \alpha\beta}\right.
\\ &\left.+\frac{1}{8}\varphi_b\left[ \frac{1}{q^2} P_T^{\alpha\beta}G_s^T(q)+\frac{1}{q^2} P_L^{\alpha\beta}G_s^L(q)\right]h_{b\ \alpha\beta}+\frac{d-2\Delta_{UV}}{2}\varphi_b G_s(q)\varphi_b\right].
\label{regaction6}
\end{align}
where
\begin{align}
&G_s^T(q)=\frac{d(d-2)}{d-1}G^T(q)+d G^L(q)+2(2d-3\Delta_{UV}),\\
&G_s^L(q)=d G^T(q)+2(2d-3\Delta_{UV}).
\end{align}
Therefore we find
\begin{eqnarray}
  G_s^T(q) &=& \frac{2\Delta_{UV}(2\Delta_{UV}-d)}{d-1}+G_s^L(q), \\
  G_s^L(q) &=& 4(d-2\Delta_{UV}),  \\
  G_s^T(q) - G_s^L(q) &=& \frac{2\Delta_{UV}(2\Delta_{UV}-d)}{d-1}.
\end{eqnarray}

The two-point function of the energy-momentum tensor can be obtained by taking the derivative of the action twice with respect to the boundary metric, and at low momentum is determined by \eqref{gtt}
\begin{equation}\label{TTcorr}
\vev{T^{\mu\nu} T^{\alpha\beta}}\propto \frac{\Lir^d}{\kappa^2\Luv}\frac{1}{(Q^2)^2}\Pi^{\mu\nu,\alpha\beta}Q^d\log(\Lir Q)^2.
\end{equation}
up to a numerical factor depending on the number of dimensions. Note also that the overall coefficient is dimensionless. This is the expected behavior when the dual field theory flows to an IR CFT. As expected for spontaneous breaking of conformal symmetry, there are no contributions to the trace of the energy-momentum tensor in the two-point function.

The scalar correlation function is obtained from \eqref{regaction6} taking the derivative twice with respect to the source $J_b$, defined in \eqref{sourceJ},
\begin{equation}\label{OOcorr}
\vev{\cO\cO}=\frac{\kappa^2\Luv}{2\Delta_{UV}-d}\vev{\cO}^2 G_s(q)\propto \frac{\kappa^2\Lir^{d+2-2\Delta_{IR}}}{\Luv}\vev{\cO}^2Q^{d-2\Delta_{IR}}.
\end{equation}
up to a numerical factor depending on the number of dimensions and the dimension of the scalar operator $\Delta_{UV}$. Note that the dimension of the two-point function agrees with its UV value $2\Delta_{UV}-d$. Since $\Delta_{IR}>d$, there is a singularity at small momentum, which is in agreement with the expectations from field theory. However, if the dilaton was a free massless mode, we would expect this singularity to be a pole, we will comment more on this in the Discussion. Note that for a generic value of $\Delta_{IR}$ a pole cannot appear at higher order in the momentum expansion because the overall singular factor will be multiplied by integer powers of $Q^2$. On the other hand, if $\Delta_{IR}$ is an integer such a pole could appear, but checking if this is the case would require a separate analysis, since the expansion we have used throughout the paper assumes that $\Delta_{IR}$ is not an integer.

The mixed correlator of the energy momentum-tensor and the scalar operator is obtained by taking a derivative with respect to each source:
\begin{equation}
\left< T^{\mu\nu} \mathcal{O} \right> = \frac{\left<\mathcal{O}\right>}{4(2\Delta_{UV}-d)} \left[ \frac{1}{q^2} P_T^{\mu\nu}G_s^T(q) + \frac{1}{q^2} P_L^{\mu\nu}G_s^L(q) \right].
\end{equation}
Writing the projection operators explicitly we are left with
\begin{align}
\notag \left< T^{\mu\nu} \mathcal{O} \right> &= \frac{\left<\mathcal{O}\right>}{4(2\Delta_{UV}-d)} \left[  \frac{1}{q^2} P_T^{\mu\nu}(G_s^T -G_s^L )+\eta ^{\mu\nu}G_s^L(q)\right]\\
\label{TOcorr} &=\frac{\Delta_{UV}\left<\mathcal{O}\right>}{2(d-1)} \frac{1}{q^2} P_T^{\mu\nu}-\vev{\cO}\eta ^{\mu\nu}.
\end{align}
The first term is the one expected when conformal symmetry is spontaneously broken. The second term is a contact term, that appears in the retarded correlator
\begin{equation}
\vev{T^{\mu\nu}(x)\cO(y)}_R\sim \vev{\cO}\eta^{\mu\nu}\delta^{(d)}(x-y),
\end{equation}
A similar term appears in the correlation function computed in the Coulomb branch flow \cite{Mueck:2001cy,Bianchi:2001de,Papadimitriou:2004rz}. In \cite{Bianchi:2001de} it was argued that this term appears because the generating functional has a term of the form $\sim \vev{\cO}_{J_b} J_b$, which gives an additional contribution. Subtracting it one would recover the true energy-momentum tensor of the field theory in the absence of sources.

We have argued that the spurion does not correspond to the Goldstone boson of spontaneous breaking of conformal invariance according to the usual holographic dictionary between normalizable solutions and dynamical modes. However, correlators of the energy-momentum tensor and scalar operator have a low energy-momentum behavior which is in agreement with the expectations from field theory. So what is the dilaton? It should correspond to a normalizable solution of the equations of motion at zero momentum. The normalizable solution in \eqref{q2sol} is the one with $C_2=0$, but the condition \eqref{cond} at $q^2=0$ implies that $C_1=0$ as well. Therefore, the only possible solution  has the same form as the linearized form of the spurion, but with appropriate boundary conditions for the metric:
\begin{equation}
\varphi=\tau(x)\partial W, \ \ h_{\mu\nu}=-2\tau(x)(W-W(0))\eta_{\mu\nu}.
\end{equation}
Note the similarity with the linearized form of the spurion, the difference is in the boundary value of the metric, that for the dilaton mode is fixed while for the spurion it changes by a Weyl factor.

\subsection{Comparison with field theory}\label{sec:compare}

Let us first show that the results we have obtained  are consistent with low-energy theorems for a field theory with spontaneous breaking of symmetry. There are two versions of the theorem, that can be found in textbooks, e.g. \cite{Weinberg:1996kr} and that we will sketch again here for convenience. The first version is based on the effective action. Suppose the action and measure of the path integral are invariant under a continuous symmetry, and for simplicity let us assume translation invariance and constant field configurations
\begin{equation}
\delta \phi_m =i \epsilon t_{mn}\phi_n,
\end{equation}
where $\phi_n$ are generic fields, $\epsilon$ a small parameter and $t_{mn}$ the symmetry generators.
If the transformation is a dilatation then $t_{mn}$ is the matrix of conformal dimensions.
The effective potential will  be invariant under this symmetry
\begin{equation}
\sum_{m,n}\frac{\delta V[\phi]}{\delta \phi_m} t_{mn} \phi_n =0.
\end{equation}
Taking a derivative of this expression with respect to $\phi_l$, one finds the following condition
\begin{equation}
\sum_m \frac{\delta V[\phi]}{\delta \phi_m} t_{ml} +\sum_{m,n}\frac{\delta^2 V[\phi]}{\delta \phi_m\delta \phi_l} t_{mn} \phi_n =0.
\end{equation}
In a vacuum configuration, the first term vanishes since it corresponds to a minimum of the effective potential. The second term can be identified with the inverse of the zero-momentum propagator
\begin{equation}
\sum_{m,n}\Delta^{-1}_{ml}(0)  t_{mn} \phi_n =0.
\end{equation}
Therefore, there is an eigenvector of the zero-momentum inverse propagator with zero eigenvalue. This is  usually interpreted as a pole at zero momentum for the propagator itself. However, from this argument one can only establish the existence of a singularity at zero momentum, not the nature of the singularity itself. The result we find for \eqref{OOcorr} is then consistent with this version of Goldstone's theorem.

The second version of the theorem is based on the Ward identities of the theory. For concreteness we will specialize to space-time transformations. Using the energy-momentum tensor we can construct the following charges that generate translations, Lorentz transformations and dilatations:
\begin{equation}
P^\mu=\int d^3 x\, T^{\mu 0}, \ \ M^{\mu\nu}=\int d^3 x\, x^{[\mu}T^{\nu]0}, \ \ D^\mu=\int d^3 x\, x_\nu T^{\nu 0}.
\end{equation}
When conformal symmetry is spontaneously broken by a constant expectation value of an operator $\cO$ of dimension $\Delta$, the commutator of the generators with the operator vanish for translations and Lorentz transformations, but not for dilatations.
\begin{equation}\label{comm}
\vev{[P^\mu,\cO]}=0, \ \ \vev{[M^{\mu\nu},\cO]}=0, \ \ \vev{[D^\mu,\cO]}=-\Delta \vev{\cO}.
\end{equation}
On the other hand, since the symmetry is only spontaneously broken, the currents associated to the symmetries should be conserved
\begin{equation}\label{ward}
\partial_\nu^y\vev{[T^{\mu\nu}(y),\cO(x)]}=0, \ \ \partial_\alpha^y\vev{[y^{[\mu}T^{\nu]\alpha}(y),\cO(x)]}=0, \ \ \partial_\mu^y\vev{[y_\nu T^{\nu\mu}(y),\cO(x)]}=0.
\end{equation}
The next step is to write the commutator between the energy-momentum tensor and the scalar field in terms of a spectral decomposition:
\begin{equation}
\vev{[T^{\mu\nu}(y),\cO(x)]}=\int \frac{d^p}{(2\pi)^3}\left[\rho^{\mu\nu}(p)e^{ip(y-x)}+\overline{\rho}^{\mu\nu}(p)e^{-ip(y-x)} \right].
\end{equation}
From \eqref{comm} and \eqref{ward}, one finds that the spectral function should take the form
\begin{equation}
\rho^{\mu\nu}(p)=(\eta^{\mu\nu}p^2-p^\mu p^\nu)\theta(p_0)\rho(-p^2),
\end{equation}
with the condition
\begin{equation}
\mu^2 \rho(\mu^2)=0, \ \ \int d\mu^2\, \rho(\mu^2)=\Delta \vev{\cO}.
\end{equation}
The two conditions can be satisfied only if
\begin{equation}
\rho(\mu^2)=\Delta \vev{\cO}\delta(\mu^2).
\end{equation}
A delta function in the commutator between $T^{\mu\nu}$ and $\cO$ implies that there should be a pole in the real part of the time-ordered correlators, proportional to $\Delta \vev{\cO}$. The holographic calculation \eqref{TOcorr} indeed shows it.

\subsection{Promoting the spurion to a dynamical mode}\label{sec:promote}

Since the spurion behaves as a regular and normalizable mode for the scalar field, we can promote it to a dynamical mode by introducing a new set of boundary conditions for the metric. This may be interesting in the context of effective low energy theories for applications to condensed matter, as it can be an example of ``holographic deconstruction'' following the proposal made in \cite{Nickel:2010pr}. In this case, the interesting IR dynamics is captured by the geometry below some cutoff in the radial direction. The IR cutoff effectively gaps the modes living between the cutoff and the boundary, but the information about the UV physics enters through the coupling of the modes below the cutoff with a massless mode, a ``Goldstone boson'' with a radial profile between the IR cutoff and the boundary. The spurion field will be one of these modes for theories with spontaneous breaking of conformal invariance.

Starting with the on-shell action \eqref{action} we can impose Dirichlet conditions for the scalar field (we will fix the source to vanish, so this is allowed) and a mixed condition for the metric if we add additional boundary terms. First, we split the variation into the traceless and the trace contributions
\begin{equation}
\delta G_{\mu\nu}=G_{\mu\nu}\delta \tau(x)+\delta G_{\mu\nu}^T, \ \ G^{\mu\nu}\delta G_{\mu\nu}^T=0.
\end{equation}
We will impose a Dirichlet condition as usual for the traceless part $\delta G_{\mu\nu}^T=0$,\footnote{Although the Weyl transformation also changes the traceless components, we are ultimately interested in the effective action of the dilaton in flat space, where these components vanish and the Weyl transformation does not affect them.} but for the trace part we will introduce a Neumann condition. In order to have a good variational principle we should add boundary terms to the action that cancel the contribution from the on-shell action $S_{\rm tot}=S+I_{UV}+I_{IR}$
\begin{equation}\label{bc}
I_{UV}=\frac{d-1}{2\kappa^2\rhuv}\int d^d x \sqrt{-\tilde G(\rhuv)}, \ \ I_{IR}=-\frac{d-1}{2\kappa^2\rhir}\int d^d x \sqrt{-\tilde G(\rhir)}.
\end{equation}
Note that $\sqrt{-G}$ in \eqref{variation} has an additional factor $\ell/2\rho$.

Now, using \eqref{variation}, the total variation of the action is
\begin{equation}
\delta S_{\rm tot}[\rhuv,\rhir]=\delta S[\rhuv,\rhir]+\delta I_{UV}+\delta I_{IR}=0.
\end{equation}
Imposing the Dirichlet conditions $\delta\phi=0$,  $\delta G_{\mu\nu}^T=0$, this takes the form
\begin{equation}\label{offhellvariation}
\delta S_{\rm tot}[\rhuv,\rhir]=-\left.\frac{1}{2\kappa^2}\int d^d x\,\sqrt{-G} \left(G_{\mu\nu} \pi^{\mu\nu}-\frac{d(d-1)}{\ell} \right)\delta\tau(x)\right|^{\rhuv}_{\rhir}.
\end{equation}
The variation vanish for any $\delta \tau(x)$ if
\begin{equation}\label{trmetbc}
\left. g^{\alpha\beta}\rho \partial_\rho g_{\alpha\beta}\right|^\rhuv_\rhir=0
\end{equation}
Since the variation is evaluated at the cutoffs, we can use the expansion \eqref{metexp} in the boundary condition \eqref{trmetbc}. At the boundary there are divergent terms that should be removed with local counterterms. The reminder in the $\rhuv\to 0$ limit is a finite contribution. For $d>2$, the leading contribution in the derivative expansion of \eqref{trmetbc}  is
\begin{equation}
\rhir \,g^{(0)\, \alpha\beta}g_{\alpha\beta}^{(2)}+\cdots =0.
\end{equation}
This will determine the equations of motion of the dilaton. For a flat background metric $g_{\mu\nu}^{(0)}=\eta_{\mu\nu}$, the infinitesimal PBH transformation \eqref{inftPBH} changes the metric to
\begin{equation}
g_{\mu\nu}^{(2)}\simeq \Lir^2\partial_\mu\partial_\nu\delta\tau,
\end{equation}
so, to leading order, the boundary condition \eqref{trmetbc} is satisfied for a massless mode
\begin{equation}
\square \delta\tau\simeq 0.
\end{equation}
The full equations of motion can be derived from the effective action of the dilaton, that can be constructed by computing the generating functional as before including the boundary contributions \eqref{bc} and performing the non-linear PBH transformation. The anomalous term with a coefficient proportional to the difference of central charges will also appear in this case.

\section{Discussion}\label{sec:discuss}

We have shown how the conformal anomaly matching arguments presented in \cite{Komargodski:2011vj,Komargodski:2011xv} are realized in theories with holographic duals. The dilaton/spurion is introduced through a coordinate transformation and our analysis shows that the effective action contains a term whose  variation is the difference between the anomalies of the UV and IR theory.
$$
\delta_\sigma \Gamma \propto  \sigma \left(\cA_d^{UV}-\cA_d^{IR}\right).
$$
The term is naturally present in even dimensions and absent in odd dimensions, where there is no anomaly.

Let us recapture the various assumptions that we made. The analysis we have presented is valid for theories whose holographic dual consists of Einstein gravity plus matter fields, with a geometry that interpolates smoothly two $AdS$ spaces in the UV and IR.
We have assumed that the potential is derived from a global superpotential which is a somewhat mild assumption in view of the fact that it allows for a quite general form of a potential. We have assumed that the scalar operator $\mathcal{O}$ carries a non-integer dimension. We made this assumption to facilitate the tedious analysis which goes into finding the small momentum approximate solution. Tacitly we are assuming that the usual AdS/CFT correspondence holds, i.e. a large N (even though we do not identify one CFT, hence the role of N in this underlying theory is not apparent) and weak-strong duality. This allows us to identify the on-shell action with classical gravitational action evaluated on the solution. Moreover we do not include higher curvature gravity terms, hence we consider only theories with $a=c$.
The null energy condition guarantees $a_{UV}\geq a_{IR}$ and makes the coefficient of the anomalous term positive definite. This is a general statement within the present, best understood, class of holographic models, however it is not quite close to be a general statement about field theories (even leaving large-$N$ and strong coupling aside). Note, in the first place, that in the holographic theories we are considering the coefficients of the anomaly polynomial at the fixed points are not independent, but there is a single independent coefficient which can be taken to be the coefficient of the Euler density. For instance, in four dimensions
\begin{equation}
\cA_4= c\cW^2-a E_4,
\end{equation}
$c=a$ is fixed. In particular this means that the holographic $c$-theorem (or rather $a$-theorem) does not apply only to the $a$ coefficient but to all the other coefficients of the anomaly polynomial. We know this to be not the case in general (e.g. \cite{Anselmi:1997am,Anselmi:1997ys}). It will then be interesting to extend the analysis to cases where $a\neq c$. In holographic models this can be achieved by including higher derivative curvature corrections, an example is Gauss-Bonnet gravity in five dimensions, for which a holographic $a$-theorem also exists for the $a$ coefficient \cite{Myers:2010xs}. An interesting observation is that the relation between the anomaly coefficient and the radius of anti-de Sitter is not straightforward when higher derivative corrections are included, so that flows where the radius of anti-de Sitter is larger in the dual to the IR theory could be allowed.\footnote{We thank Andrei Parnachev for pointing this out to us.}

Let us now comment on the behavior of the scalar correlator for the case of spontaneous breaking, whose low-momentum behavior we have found to be
$$
\vev{\cO\cO}\sim (\sqrt{-q^2})^{d-2\Delta_{IR}}.
$$
where $\Delta_{IR}>d$. Note that this is more singular than the pole behavior we expected to see.  We can compare our results with the results for the singular Coulomb branch geometry. There, it was found that there is a pole in this correlator \cite{Mueck:2001cy,Bianchi:2001de,Papadimitriou:2004rz}. Although we understand this as coming from the properties of the solution close to the horizon, we are not sure about the physical origin of this difference in the two cases. A qualitative difference is that in the Coulomb branch the spectrum is gapped, which probably limits the singular behavior of the scalar correlator to a pole, while in the cases we study there is always a large number of massless degrees of freedom. It would be interesting to explore other models with spontaneous breaking of global symmetries, and compare the two-point function of the operator that triggers the breaking in cases where the spectrum is gapped with those where there are more massless states beside the Goldstone bosons.

There are several examples where massless degrees of freedom affect the qualitative behavior of correlators. For instance, in the Landau-Ginzburg model long-range interactions can in general modify the momentum dependence of the correlation function of the order parameter, leading to non-analytic dependence on momentum \cite{Ferrell:1972zz}. Long-range interactions can also modify the dispersion relation of Goldstone bosons in non-relativistic theories (some examples are discussed in section 6 of \cite{Brauner:2010wm}), avoiding the conclusions of the Nielsen-Chadha theorem \cite{Nielsen:1975hm}.

Said this, the behavior of the scalar correlator is quite surprising to us. Our na\"ive expectation was that $\cO$ would reproduce the propagator of a weakly coupled dilaton, that would lead to a different momentum dependence. Assume that, as na\"ively expected, the low energy effective field theory description consists of the IR CFT plus an almost free dilaton. The dilaton does not couple directly to marginal operators, so the only allowed couplings to the CFT are to irrelevant operators, that would decouple at very low energy. The effective action at low energies in that case would be
$$
\cL_{IR}\simeq \cL_{IR\, CFT}+\cL_{\rm dilaton}+g\left(\vev{\cO}\right)e^{\Delta_{IR}\tau} \cO_{\Delta_{IR}}.
$$
Here $g(\vev{\cO})$ is the coupling of the irrelevant operator that drives the IR theory away from the fixed point in the RG flow. This coupling depends on the expectation value of the operator that breaks conformal invariance. Using that $\vev{\cO_{IR}\cO_{IR}}\sim (\sqrt{-q^2})^{2\Delta_{IR}-d}$, the dilaton propagator would take the form
$$
\vev{\tau \tau}\sim \frac{1}{q^2+(\sqrt{-q^2})^{2\Delta_{IR}-d}}.
$$
Clearly, this is different from the scalar correlator $\vev{\cO\cO}$ we have computed. In our case, the low-momentum behavior of the correlator coincides with that of an operator of dimension $d-\Delta_{IR}$ in the IR CFT.
$$
\vev{\cO\cO}\sim (\sqrt{-q^2})^{2(d-\Delta_{IR})-d}.
$$
This is the dimension of the {\em coupling} $g(\vev{\cO})$. We do not have a formal argument in field theory that explains this behavior, heuristically it looks like fluctuations of the expectation value lead to fluctuations of the coupling in the effective IR theory. Note that there is a strong IR divergence, the space-dependent Euclidean correlator grows like a power-law at long distances:
$$
\vev{\cO(x)\cO(0)}\sim |x|^{2(\Delta_{IR}-d)}.
$$
This is even stronger than the logarithmic divergence of a massless scalar field in two dimensions, that precludes the formation of a condensate \cite{Coleman:1973ci}. Taken at face value, it implies that there is no unitary field theory dual to the class of models we have chosen. This behavior may be avoided in different models with no global definition of a superpotential or with more than one scalar field. Recall that having a globally defined superpotential is not a generic situation. The power law we observe in the scalar correlator seems to be related to the coefficient of the quadratic term of the superpotential at the IR critical value $\phi_m$, which for a single scalar field in the bulk is necessarily negative. However, when there are more scalar fields classical trajectories in field space could connect two critical points of the superpotential along directions of positive curvature, a cartoon is given in figure \ref{fig1}. Another possibility is that there is no global definition of the superpotential, so close to the IR critical value $\phi_m$ the local form of the superpotential has a positive coefficient in front of the quadratic term.

\FIGURE[h]{
  \includegraphics[width=8.5cm]{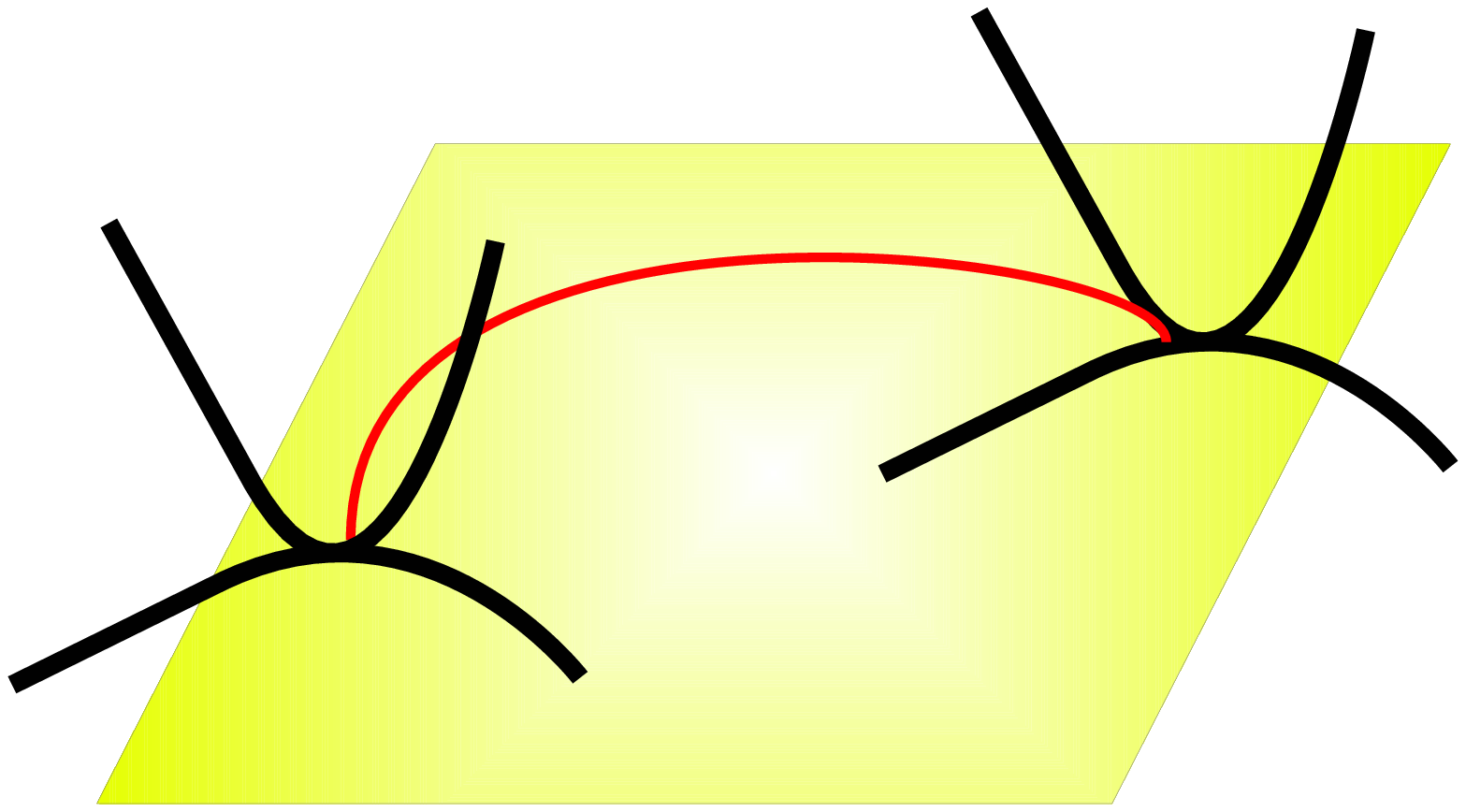}
	\caption{ Cartoon of a trajectory between two critical points of the superpotential in a two-dimensional space. The trajectory approaches the critical points along the directions of positive curvature.
	}
	\label{fig1}
}

One may ask whether the model we constructed really corresponds to spontaneous breaking of symmetry. We already made some arguments in the Introduction that this is indeed the case, let us elaborate on them. In the first place we have chosen the superpotential in such a way that in the background solution only the normalizable mode of the scalar is turned on, which in the field theory side is interpreted as having a non-zero expectation value for the dual operator but no sources. Since there are no sources, the renormalized action that we have computed is independent of the expectation value.\footnote{Note that this may be different for marginal operators, or operators with integer dimensions such that multi-trace operators are marginal.} Therefore, we are allowed to make the bulk transformation
\begin{equation}
\delta\phi=\tau(x)\partial W, \ \ \delta g_{\mu\nu}=-2\tau(x)(W-W(0))\eta_{\mu\nu}.
\end{equation}
which shifts the coefficient of the normalizable scalar solution and hence of the expectation value of the dual operator. This also corresponds to the zero mode which we identified in the spectrum of fluctuations. The metric changes in the bulk, but the boundary metric is held fix. This transformation can be interpreted as having a moduli space of vacua spanned by the expectation value, with the `dilaton' mode corresponding to fluctuations along the moduli. Moreover, the boundary conditions in both the UV and IR do not break the scale invariance explicitly. In the model we study these fluctuations are very strong which leads to the singular behavior that we observe in the scalar correlator. Further evidence in support of a scenario with spontaneous breaking of symmetry is that the results for the correlation functions of the energy-momentum tensor agree with the expectations from Goldstone's theorem.

It would be interesting to study more general models with one or several scalar fields and see if the singular behavior can be avoided. These models are also interesting for other reasons, for instance they can also be used to study the spontaneous breaking of Abelian or non-Abelian global symmetries. A particularly intriguing fact is that, to our knowledge, there are no examples in consistent truncations of ten-dimensional supergravity of geometries interpolating between two AdS spaces where the flow is dual to a spontaneously broken theory. The known models are either singular or involve explicit breaking. In ten-dimensional supergravity those models can exist, in the case of $\cN=4$ SYM in the Coulomb branch one expects to have several $AdS_5$ throats corresponding to different stacks of D3 branes, and explicit examples have been constructed for $\cN=4$ itself \cite{Costa:1999sk,Costa:2000gk} and in geometries that explore the baryonic branch of Klebanov-Witten theories, interpolating between two $AdS_5$ spaces \cite{Klebanov:2007us,Martelli:2007mk}. The fact that the smooth geometries correspond to localized branes in the internal space while smooth distributions seem to lead generically to singular geometries \cite{PandoZayas:2000sq,Freedman:1999gk} suggests that the singularity is an artifact of the approximation, and that by resolving at small enough distances, the singularity will split in multiple $AdS_5$ throats.

Summarizing, our result for the correlators leads to the following possibilities:

\begin{itemize}
\item For some reason the low momentum expansion breaks down and there are additional terms in the scalar correlator that makes it less singular, modifying it to a $1/q^2$ pole. We checked corrections to the next few orders and did several consistency checks and could not figure out how this would happen. We also checked the method in the Coulomb branch singular solutions where the analytic solutions are known and found no problem there with the low momentum expansion, we reproduce previous results that found a $1/q^2$ pole in the scalar correlator. In any case, it would still be desirable to have an analytic example where the horizon geometry is $AdS$ and the low momentum expansion is not necessary.
\item The singular behavior is an artifact of the large-$N$, strong coupling approximation that is implicit in the holographic approach and a proper treatment will avoid these problems. This would imply that the supergravity approximation breaks down, which is difficult to understand from the bulk perspective, since the classical solution is completely smooth and the curvature is small everywhere.
\item The model does not correspond to any known consistent truncation of string theory, so the that the field theory dual is not known. It is possible that it does not exist, at least in the form of a quantum field theory satisfying the usual properties of locality and unitarity. One would expect that inconsistencies in the field theory would also be manifested in some form in the gravity dual. We do not observe them at the level at which we are doing our analysis, but they may appear elsewhere.
\end{itemize}

In our opinion, the most likely explanation is the last point. From the field theory point of view, in order to have spontaneous breaking of scale invariance it is necessary to have a moduli space of flat directions. These are ubiquitous in supersymmetric theories. For instance, in four dimensions the simplest example is $\cN=4$ SYM, where quantum corrections do not modify the classical moduli space. In a string theory setup, the moduli space of $\cN=4$ can be explored by moving D3 branes around. As we have explained we do not expect a fully smooth gravity dual solution describing the whole theory to the far IR when we do a consistent truncation to five dimensions, since there are always some missing low energy degrees of freedom (corresponding to the unbroken gauge groups on the D3 branes). It is probably not easy to find a situation where this difficulty is overcome. A lesson is that one should be cautions when working with holographic toy models, even when the background solutions have no obvious singularities or instabilities, other kind of inconsistencies may arise in correlation functions.
We would like to explore these issues in the future.

\section*{Acknowledgements}
We thank Ofer Aharony for many useful discussions and suggestions and Stefano Cremonesi, Ori Ganor, Rob Myers, Andrei Parnachev, Koenraad Schalm, Adam Schwimmer and Stefan Theisen for useful comments.
We also thank Ioannis Papadimitriou and Kostas Skenderis for very helpful comments on the holographic renormalization and the fluctuation analysis.
This work was supported in part by the Israel Science Foundation (grant number 1468/06).

\appendix

\section{Equations of motion of fluctuations}\label{app:eoms}

It is convenient to write the equations of motion \eqref{einsteqs} in terms of the extrinsic curvature
\begin{equation}
K_{\mu\nu}=-\Gamma^r_{\mu\nu}=\frac{1}{2}\partial_r g_{\mu\nu}, \ \ K^\mu_{\ \nu}=g^{\mu\alpha}K_{\alpha\nu}=\Gamma^\mu_{r\nu}, \ \ K=g^{\mu\nu}K_{\mu\nu}.
\end{equation}
The Einstein equations take the form (primes denote radial derivatives)
\begin{align}
&\label{eq1b}K'+K^\alpha_{\ \beta} K^\beta_{\ \alpha}=-(\phi')^2-\frac{2}{d-1} V,\\
&\label{eq2b}\partial_\mu K-\hat\nabla_\alpha K^\alpha_{\ \mu}=-\phi'\partial_\mu\phi,\\
&\label{eq3b}\hat R_{\mu\nu}+K_{\mu\nu}'-2K_{\mu\alpha}K^\alpha_{\ \nu}+K K_{\mu\nu}=-\partial_\mu\phi\partial_\nu\phi-\frac{2}{d-1}g_{\mu\nu} V,\\
&\label{eq4b}\phi''+K\phi'+g^{\mu\nu}\hat \nabla_\mu\partial_\nu\phi-\partial_\phi V=0,
\end{align}
where hats denote covariant derivatives with respect to $g_{\mu\nu}$. It is convenient to choose a slightly different set of equations. For this, we will use the formulas
\begin{align}
& K'=(g^{\mu\nu}K_{\mu\nu})'=g^{\mu\nu}K_{\mu\nu}'-2K_\mu^{\ \alpha} K_{\alpha\nu},\\
& K_{\mu\nu}'=(g_{\mu\alpha}K^\alpha_{\ \nu})'=g_{\mu\alpha}{K^\alpha_{\ \nu}}'+2 K_{\mu\alpha} K^{\alpha\nu}.
\end{align}
Now we take the trace of \eqref{eq3b} and subtract \eqref{eq1b} to get a new equation with only one radial derivative. Our last equation will be \eqref{eq3b}, but with one index raised
\begin{align}
&\label{eq1c} K^\mu_{\ \nu} K^\nu_{\ \mu}-K^2+(\phi')^2=2 V+\hat R+ g^{\mu\nu}\partial_\mu\phi\partial_\nu\phi,\\
&\label{eq2c}\partial_\mu K-\hat\nabla_\alpha K^\alpha_{\ \mu}=-\phi'\partial_\mu\phi,\\
&\label{eq3c} (K^{\mu}_{\ \nu})'+K K^{\mu}_{\ \nu}=-\hat R^\mu_{\ \nu}-\partial^\mu\phi\partial_\nu\phi-\frac{2}{d-1}\delta^{\mu}_{\ \nu} V,\\
&\label{eq4c}\phi''+K\phi'+g^{\mu\nu}\hat \nabla_\mu\partial_\nu\phi-\partial_\phi V=0,
\end{align}
In this case \eqref{eq1c} and \eqref{eq3c} are constraints associated to the gauge fixing. We now expand the equations to linear order in fluctuations
\begin{equation}
g_{\mu\nu}=e^{2A}(\eta_{\mu\nu}+h_{\mu\nu}), \ \ \phi=\phi_0+\varphi.
\end{equation}
We will need the extrinsic curvature
\begin{equation}
K^\mu_{\ \nu}=A'\delta^\mu_\nu+\frac{1}{2}{h^\mu_{\nu}}', \
 K=d A'+\frac{1}{2}h',
\end{equation}
where primes denote radial derivatives, indices are raised with the flat metric and $h=h^\mu_\mu=\eta^{\mu\nu}h_{\mu\nu}$. We will also use the Ricci tensor to linear order
\begin{equation}
\hat R^\mu_{\ \nu}=\frac{1}{2}e^{-2A}\left(\square h^\mu_{\ \nu}-\partial^\mu\partial_\lambda h^\lambda_{\ \nu}-\partial_\nu\partial_\lambda h^{\mu\lambda}+\partial^\mu\partial_\nu h\ \right).
\end{equation}
We now expand in Fourier modes, each mode admits the following decomposition in transverse and longitudinal parts, traceless and not traceless, as defined in \eqref{hdec} and \eqref{qtens}. The equations of motion are
\begin{align}
\label{eqAa} &0= (d-1)A'h'+2\partial V\varphi -2\phi_0'\varphi'-q^2 e^{-2A} h_T,\\
\label{eqB} &0= q_\mu h'-q_\alpha {h^{\alpha}_{\ \mu}}'+2q_\mu\phi_0'\varphi,\\
\notag &0={h^\mu_{\ \nu}}''+d A' {h^\mu_{\ \nu}}'+\delta^\mu_\nu \left(A'h'+\frac{4}{d-1}\partial V\varphi\right)\\ \label{eqC} &-e^{-2A}(q^2\eta^{\sigma\mu}\eta^{\rho\nu} -q^\mu q_\rho\eta^{\sigma\nu}-q^\nu q_\sigma\eta^{\rho\mu}+q^\mu q^\nu\eta^{\sigma\rho} )h_{\sigma\rho},\\
\label{eqS} &0=\varphi''+dA'\varphi'-\partial^2 V\varphi-q^2 e^{-2A}\varphi+\frac{\phi_0'}{2}h'.
\end{align}
Projecting \eqref{eqB} with $P_T^{\mu\nu}$, gives the equation ${h^{TL}_{\mu\nu}}'=0$, so we can set this mode to zero. Projecting the same equation with $P_L^{\mu\nu}$ gives
\begin{equation}\label{consTa}
0=h_T'+2\phi_0'\varphi.
\end{equation}
Projecting with $\Pi^{\alpha\beta,\nu}_{ \ \  \ \mu}$ \eqref{eqC} gives a decoupled equation for the transverse traceless mode
\begin{equation}\label{eqTTa}
0={h^{TT}_{\mu\nu}}''+dA'{h^{TT}_{\mu\nu}}'-q^2e^{-2A}{h^{TT}_{\mu\nu}}.
\end{equation}
Projecting the same equation with either $P_{T \nu}^{\ \mu}$ or $P_{L \nu}^{\ \mu}$ gives the two equations
\begin{align}
\label{eqT} &0=h_T''+dA'h_T'+(d-1)A'h'+4\partial V\varphi-q^2 e^{-2A} h_T,\\
\label{eqL} &0=h_L''+dA'h_L'+A'h'+\frac{4}{d-1}\partial V\varphi-q^2 e^{-2A} h_T.
\end{align}
It turns out that not all the equations are independent. If we subtract \eqref{eqAa} from \eqref{eqT} and use \eqref{consTa} to remove the dependence on $h_T$ we find that the result vanish identically. Therefore, from \eqref{eqA}, \eqref{consTa}, \eqref{eqT} and \eqref{eqL} we only need to keep three equations. We will combine \eqref{eqAa}, \eqref{eqT} and \eqref{eqL} such that we find an equation that depends only on $h$ and $\varphi$:
\begin{equation}\label{eqtracea}
0=h''+2 A' h'+\frac{4}{d-1}\partial V \varphi+4\phi_0'\varphi'.
\end{equation}
We can determine $\varphi$ and $h$ from \eqref{eqS} and \eqref{eqtracea}, $h_T'$ from \eqref{consTa} and $h_L'=h'-h_T'$. For the additional constraint \eqref{eqAa} to hold, it is enough to check that it is satisfied at some fixed value of the radial coordinate. Indeed, taking a derivative with respect to the radial coordinate and using the other equations, including the equations for the background, one can check that the result vanishes identically.

Let us define now $h'=e^{-2A}H$, we are left with the equations
\begin{align}\label{scalareqsa}
 &0= e^{-2A}H'+\frac{4}{d-1}\partial V \varphi+4\phi_0'\varphi',
 &0=\varphi''+dA'\varphi'-\partial^2 V\varphi-q^2 e^{-2A}\varphi+\frac{\phi_0'}{2}e^{-2A}H.
\end{align}

\section{The matching procedure}\label{app:matching}

In this section we explicitly show the existence of an overlapping region between the boundary and the near-horizon solutions.

The equation for $G$ \eqref{eqG} can be written in the following form
\begin{equation}
x^2\partial^2G+\alpha (x) x \partial G +\beta (x) Q^2 x^{\frac{2}{\lambda}}G=0
\end{equation}
where $x\equiv \phi_0 - \phi_m$ ($\phi_m$ is the radial location of the horizon). The functions $\alpha(x)$ and $\beta(x)$ can be expanded around the horizon $x=0$
\begin{eqnarray}
  \alpha(x) &=& \alpha_0 +\alpha_1 x +\alpha_2 x^2 +... \\
  \beta(x) &=& \beta_0 +\beta_1 x +\beta_2 x^2+...
\end{eqnarray}
with
\begin{eqnarray}
  \alpha_0 &=& \frac{d-2}{\lambda} -1 \\
  \beta_0 &=& \left( \frac{L_{IR}}{\lambda} \right) ^2
\end{eqnarray}
and
\begin{equation}
\lambda=d-\Delta_{IR}<0
\end{equation}

Very close to the horizon we can approximate $\alpha=\alpha_0$ and $\beta=\beta_0$. With this approximation the solution is (up to an overall constant that we drop)
\begin{equation}
G_0=\left( \lambda Q x^{\frac{1}{\lambda}} \right) ^{\frac{\lambda}{2}(1-\alpha_0)} H_{-\frac{\lambda}{2}(1-\alpha_0)} \left( \sqrt{\beta_0}\lambda Q x^{\frac{1}{\lambda}} \right)
\end{equation}
where $H$ is the Hankel function. In order to carry out the matching procedure we now have to expand this solution for $Qx^{\frac{1}{\lambda}}\ll 1$. This expansion takes the form
\begin{eqnarray}
\nonumber  G_0 &=& a \left[ 1+C_2 \left(\sqrt{\beta_0}\lambda Q x^{\frac{1}{\lambda}}\right)^2 +...\right] \\
   &+&  b \left[ 1+d_1 \left(\sqrt{\beta_0}\lambda Q x^{\frac{1}{\lambda}}\right)^2 +...\right] \left(\sqrt{\beta_0}\lambda Q x^{\frac{1}{\lambda}}\right)^{\lambda(1-\alpha_0)}
\end{eqnarray}
Our goal is to check whether subleading contributions in the near-horizon expansion can give terms with the same dependence on $x$ as the leading terms in the expansion of $G_0$.

The first correction for the solution around the horizon is determined by the equation
\begin{equation}
x^2 \partial ^2 G_1 +\alpha_0 x \partial G_1 +\beta_0 Q^2 x^{\frac{2}{\lambda}}G_1 = -x \left[ \alpha_1 x\partial G_0 + \beta_1Q^2 x^{\frac{2}{\lambda}}G_0 \right]
\end{equation}
The exact solution for $G_1$ will be some complicated superposition of Hankel functions. However, we are only interested in the expansion of the solution for large values of $x$, so we can solve this equation order by order. The solution takes the form
\begin{eqnarray}
\nonumber   G_1 &=& ax \left[ \# Q^2 x^{\frac{2}{\lambda}} + \# Q^4 x^{\frac{4}{\lambda}} +... \right] \\
   &+& bx \left[ \# + \# Q^2 x^{\frac{2}{\lambda}} +... \right] \left(\sqrt{\beta_0}\lambda Q x^{\frac{1}{\lambda}}\right)^{\lambda(1-\alpha_0)}
\end{eqnarray}
where $\#$ represents numerical coefficients (with no dependence on the momenta $Q$). In a similar manner one can solve for higher orders and the result will take a similar form (with the coefficients in front of each of the series above replaced by $ax^n$ and $bx^n$ respectively, for the $n$th order).

We conclude that the solution in the vicinity of the horizon takes the form
\begin{eqnarray}
\nonumber   G &=& a\left[ 1 + C_2(x)Q^2 x^{\frac{2}{\lambda}} + c_2(x)Q^4 x^{\frac{4}{\lambda}} +... \right] \\
   &+& b\left[ d_0(x) + d_1(x)Q^2 x^{\frac{2}{\lambda}} + d_2(x)Q^4 x^{\frac{4}{\lambda}} +... \right] \left(\sqrt{\beta_0}\lambda Q x^{\frac{1}{\lambda}}\right)^{\lambda(1-\alpha_0)}
\end{eqnarray}
where each of the functions above \{$c_i,d_i$\} can be expanded as a series in $x$ and does not depend on the momenta $Q$. In particular, we find that $d_0(0)=1$.

By examining the power structure of the expansion above, one notices that some terms appear in both series. For instance, the order $\nu$ term of the second series will mix with the order $\left(\nu-\frac{d-2}{2}\right)$ of the first series. However, the leading terms in each series do not mix. Therefore, when taking the limits $x\ll1$ and $Q^2 x^{\frac{2}{\lambda}}\ll1$ only the near-horizon solution ($G_0$) will survive and there would be no contributions from higher orders. That shows that there is a region where the boundary solution \eqref{q2sol} and the near-horizon solution \eqref{horizonSol} overlap.

\section{Matching procedure for Coulomb branch solution}\label{app:coulomb}

In this section we check the matching procedure at low momentum for the Coulomb branch solutions. Our conventions are different from those in \cite{Mueck:2001cy},  both for the gauge fixing and the definition of the scalar action. In our conventions, the superpotential is
\begin{equation}
W=\frac{v+2}{v^{1/3}}, \ \ v=e^{\sqrt{3}\phi}.
\end{equation}

\subsection{Tensor fluctuation}

The equation of motion for the tensor fluctuation is the same as \eqref{eqTT}, using $v$ as variable we can also write it as
\begin{equation}\label{eqtenc}
v(v-1)\partial_v^2 h^{TT}+(v-2)\partial_v h^{TT}+\frac{Q^2}{4v} h^{TT}=0.
\end{equation}
In this coordinate, $v=0$ is the horizon (which is singular) and $v=1$ the boundary. The exact solutions are
\begin{equation}
h^{TT}=C_1 v^{-\alpha}{}_2 F_1\left[-\frac{\alpha}{2},-\frac{\alpha}{2},2-\alpha; v \right]+C_2 v^{\alpha-2}{}_2 F_1\left[\frac{\alpha}{2}-1,\frac{\alpha}{2}-1,\alpha; v \right],
\end{equation}
where $\alpha=\sqrt{1+Q^2}+1$. Imposing regularity at the horizon sets $C_1=0$ and $C_2=h_b^{TT}$. The expansion of the regular solution at low momentum takes the form
\begin{equation}\label{expansionht}
h^{TT} \simeq h_b^{TT}\left(1+\frac{Q^2}{4}\log v +\mathcal{O}(Q^4) \right).
\end{equation}
So $h_b^{TT}$ can be identified as the source of the energy-momentum tensor up to kinematical factors.

Let us now see how this behavior is captured by the matching procedure. First, we take the near-horizon limit $v\to 0$, where the equation of motion \eqref{eqtenc} takes the form
\begin{equation}
\partial_v^2 h^{TT}+\frac{2}{v}\partial_v h^{TT}-\frac{Q^2}{4v^2}h^{TT}=0.
\end{equation}
The solutions to this equation are
\begin{equation}
h^{TT}=A_0 v^{\frac{\alpha}{2}-1}+A_1 v^{-\frac{\alpha}{2}}.
\end{equation}
Expanding at low momentum $Q^2\log v \ll 1$ gives
\begin{equation}
h^{TT}\simeq A_0\left[1+\frac{Q^2}{4}\log v\right]+\frac{A_1}{v}\left[1-\frac{Q^2}{4}\log v\right] +\mathcal{O}\left(\frac{Q^4}{4}(\log v)^2 \right).
\end{equation}

We now consider the $Q^2\to 0$ limit, where the solution has the form \eqref{tenssol}, and expand for small $v$
\begin{eqnarray}
\nonumber h^{TT} &\simeq& h_b^{TT}+h_{N}^{TT}\int d\phi_0\,\frac{1}{\phi_0'}\, e^{-4A}+\mathcal{O}(Q^2) \\
\nonumber &=& h_b^{TT}+h_{N}^{TT}\int \frac{dv}{(\partial v/\partial \phi)}\frac{1}{\phi_0'}e^{-4A} +\mathcal{O}(Q^2) \\
&\simeq& h_b^{TT}-\frac{1}{2} h_{N}^{TT}\left(\frac{1}{v}+\log v\right)+\mathcal{O}(Q^2)
\end{eqnarray}
Including the correction to next order we have
\begin{equation}
h^{TT}\simeq h_b^{TT}+\left(-\frac{1}{2}h_{N}^{TT}+\frac{Q^2}{4}(h_{N}^{TT}-h_b^{TT}) \right)\frac{1}{v}-\frac{1}{2}h_{N}^{TT}\left(1-\frac{Q^2}{4 v} \right)\log v.
\end{equation}
We can match the two expressions if
\begin{equation}
A_0=h_b^{TT}, A_1=-\frac{1}{2}h_{N}^{TT}+\frac{Q^2}{4}(h_{N}^{TT}-h_b^{TT}),
\end{equation}
and $h_{N}^{TT}\propto Q^2$ so the expansion is well defined. The regularity condition $A_1=0$ implies
\begin{equation}
h_{N}^{TT}=-\frac{Q^2}{2} h_b^{TT},
\end{equation}
to leading order in the expansion. Therefore, the solution becomes
\begin{eqnarray}
\nonumber h^{TT} &\simeq& h_b^{TT}\left(1+\frac{Q^2}{4}\log v+\mathcal{O}(Q^4) \right)
\end{eqnarray}
in agreement with the expansion of the exact solution \eqref{expansionht}.

\subsection{Scalar fluctuation}

The equation of motion for the scalar fluctuation is given by \eqref{eqscalar}. In terms of the $v$ coordinate it is given by
\begin{eqnarray}\label{eq3rdorder}
\nonumber \partial_v^3\varphi+\frac{13-10 v}{3 v-3 v^2}\partial_v^2\varphi &+\frac{3 Q^2 (v-1)+8 \left(v^2-2 v+4\right)}{12 (v-1)^2 v^2}\partial_v\varphi \\
&+\frac{Q^2 \left(-2 v^2+v+1\right)-24 v}{12 (v-1)^3 v^3}\varphi &=0.
\end{eqnarray}
Close to the boundary $v\to 1$ there is a regular singular point, so this equation has three independent solutions
\begin{equation}
\varphi\simeq a_0 (v-1)+a_1 (v-1)\log(1-v)+\left(a_2- \frac{a_0}{3}\right) (v-1)^2.
\end{equation}
The solution proportional to $a_0$ has the asymptotic behavior of the gauge solution $\varphi =\partial W$, that is an exact solution to the equations of motion. The other two solutions can be obtained from the computation of the gauge-invariant combination \eqref{gaugeinvG}, that obeys the equation \eqref{eqG}. In terms of the $v$ coordinate, it becomes
\begin{equation}\label{eqG2c}
\partial_v^2 G+\frac{2}{v+2}\partial_v G+\frac{Q^2}{4 (v-1) v^2}G=0.
\end{equation}
The exact solutions are
\begin{equation}
G=B_0 \frac{1}{v+2}v^{1-\alpha/2}{}_2 F_1\left[-\frac{\alpha}{2},-\frac{\alpha}{2},2-\alpha; v \right]+B_1 \frac{1}{v+2}v^{\alpha/2}{}_2 F_1\left[\frac{\alpha}{2}-1,\frac{\alpha}{2},\alpha; v \right],
\end{equation}
where $\alpha=\sqrt{1+Q^2}+1$. Imposing regularity at the horizon sets $B_0=0$.
For the regular solution, the expansion close to the boundary at low momentum is, to leading order
\begin{equation}\label{exact}
\frac{G}{B_1} \simeq \frac{1}{3} +(v-1)\left[\frac{2}{9}-\frac{1}{12}Q^2\log(1-v)\right].
\end{equation}

Let us now show how the matching works. First, we approximate \eqref{eqG2c} by the near-horizon form
\begin{equation}
\partial_v^2 G-\frac{Q^2}{4 v^2}G=0.
\end{equation}
Solutions to this equation are
\begin{equation}
G\simeq A_0 v^{1-\alpha/2}+A_1 v^{\alpha/2},
\end{equation}
the regularity condition being $A_0=0$. Expanding for low momentum gives
\begin{equation}\label{horexp}
G\simeq A_1 v\left(1+\frac{Q^2}{4}\log v \right).
\end{equation}

We now use the solution in the low momentum limit \eqref{q2sol}
\begin{align}\label{boundexp}
\notag  G \simeq C_2+\frac{C_1}{v+2}&-\frac{Q^2}{4} C_2\left(\frac{ 2 + 3 (v-1 ) \log(1 - v) + (2 - 3 v) \log v }{4 ( v+2)} \right)\\ &-\frac{Q^2}{4}C_1\left( \frac{(v-1) \log (1-v)-v \log v+\log v+1}{v+2}\right).
\end{align}
Expanding close to the horizon, one gets
\begin{align}
 G &\simeq \left(C_2+\frac{C_1}{2}\right)\left(1-\frac{1}{4}Q^2\log v\right) +v\left(-\frac{C_1}{4}+\frac{8C_2+3C_1}{16}Q^2\log v \right).
\end{align}
Setting $C_2=-C_1/2$ and $C_1=-4 A_1$ matches with \eqref{horexp}
\begin{align}
\notag G &\simeq A_1 v\left(1+\frac{Q^2}{4}\log v \right).
\end{align}
Expanding now \eqref{boundexp} close to the boundary,
\begin{equation}
\frac{G}{2A_1}\simeq \frac{1}{3}+(v-1)\left[\frac{2}{9}-\frac{Q^2}{12}\log(1-v)\right].
\end{equation}
This is in agreement with \eqref{exact}, therefore the matching procedure fixes the leading terms in the low momentum expansion.

\section{RG flow toy models}\label{app:toy}

In order to have an $AdS$ metric when $\phi\to \phi_m$, it is necessary that $W'(\phi)\sim (\phi-\phi_m)$, so the superpotential has a critical point $W'(\phi_m)=0$.\footnote{Notice that critical points of the superpotential are always critical points of the potential, since $V'(\phi)=W'(\phi)\left( W''(\phi)-\frac{d}{d-1}W(\phi)\right)$.} Then, in order to have a soliton solution between $\phi=0$ and $\phi=\phi_m$, we will require that the superpotential satisfies the following conditions (for $\phi_m>0$ and $\delta=\Luv/\Lir>1$)
\begin{align}\label{eq:supcond}
&W(0)=\frac{d-1}{\Luv}, \ \ W'(0)=0, \ \ W''(0)=\frac{\Delta_{UV}}{\Luv}>0,\\
&W(\phi_0)=\frac{d-1}{\Lir}, \ \ W'(\phi_0)=0, \ \ W''(\phi_0)=\frac{d-\Delta_{IR}}{\Lir}<0.
\end{align}
There are many such possible superpotentials, so we need some additional criteria to make our choice. In supergravity theories this kind of superpotentials can be found after doing a dimensional reduction in some internal manifold (the scalars correspond to the moduli) and are usually a sum of exponentials. Inspired by this fact, we will make our superpotential also a sum of exponentials. In order to find two critical points we need at least three exponential terms, a possible ``minimal'' superpotential is then
\begin{equation}
W(\phi)=W_0\left(e^{a \phi}+\frac{e^{-a \phi}}{1+k+k^2}-\frac{(k+k^2)}{2(1+k+k^2)}e^{2 a \phi} \right).
\end{equation}
By construction, $W(\phi)$ has two critical points, one at $\phi=0$ and the other at $\phi=\phi_m=-1/a \log(k)$.

Using the scalar field as radial coordinate,
\begin{equation}
ds^2=g_{\phi\phi}d\phi^2+e^{2A(\phi)}\eta_{\mu\nu}dx^\mu dx^\nu,
\end{equation}
the coefficients of the metric are
\begin{equation}
g_{\phi\phi}=\frac{1}{\left(W'(\phi)\right)^2}=\frac{1}{a^2 W_0^2}\left(e^{a \phi}-\frac{e^{-a \phi}}{1+k+k^2}-2\frac{(k+k^2)}{2(1+k+k^2)}e^{2 a \phi} \right)^{-2},
\end{equation}
and
\begin{align}
\notag A(\phi) &=A_0+\frac{\phi}{a(d-1)}-\frac{1}{2 a^2(d-1)}\Big[\frac{4+k+k^2}{2-k-k^2} \log\left(e^{a\phi}-1\right)\\ &-\frac{4+k+k^2}{(1-k)(2k+1)} \log\left(1-k e^{a\phi}\right)+\frac{4+7k+4k^2}{(2+k)(1+ 2 k)}\log\left(1+(k+1) e^{a\phi}\right)\Big],
\end{align}
where $A_0$ is an arbitrary integration constant. The radial coordinate is
\begin{align}
\notag r(\phi)&=-\int^\phi \frac{d\tilde\phi}{W'(\tilde\phi)}= r_0-\frac{(1+k+k^2)}{a^2 W_0(2-k-k^2)}\Big[ \log\left(e^{a\phi}-1\right)-\frac{k(k+1)}{(2k+1)} \log\left(1-k e^{a\phi}\right)\\ &-\frac{1-k^2}{(1+ 2 k)}\log\left(1+(k+1) e^{a\phi}\right)\Big],
\end{align}
where $r_0$ is the integration constant.

One can satisfy all the conditions \eqref{eq:supcond} if
\begin{equation}
W_0=2\frac{d-1}{\Luv}\frac{1+k+k^2}{4+k+k^2}, \ \ a=\sqrt{\frac{4+k+k^2}{2-k-k^2}}\sqrt{\frac{\Delta_{UV}}{2(d-1)}}, \ \ 0<k<1.
\end{equation}
Close to $\phi=0$,
\begin{align}
&r \sim -\frac{\Luv}{\Delta_{UV}} \log(\phi) \ \ \Rightarrow \ \ \phi \sim e^{-\Delta_{UV} r/\Luv}, \ r\to \infty,\\
&A\sim -\frac{1}{\Delta_{UV}} \log(\phi) \ \ \Rightarrow \ \ e^{2 A} \sim e^{2 r/\Luv}.
\end{align}
This shows that the metric indeed becomes $AdS$. Furthermore, the profile of the scalar field corresponds to the normalizable solution, which means that in the dual field theory the scalar operator has a vacuum expectation value but there is no explicit breaking of conformal invariance.

The ratio between the UV and IR $AdS$ radii is
\begin{equation}
\frac{\Lir}{\Luv}=\frac{1}{\delta}=\frac{k(4+k+k^2)}{1+k+4 k^2}.
\end{equation}
The IR dimension is
\begin{equation}
\Delta_{IR}=d-\Lir W''(\phi_m)=d+\Delta_{UV}\frac{(1+2k)(4+k+k^2)}{(2+k)(1+k+4 k^2)}.
\end{equation}
We have constructed a one-family parameter of soliton solutions for any given $d$ and $d>\Delta_{UV}>0$.

\end{document}